\documentclass[preprint,review,12pt]{elsarticle}
\usepackage{amssymb}
\usepackage{graphicx}
\journal{Nuclear Physics A}

\def\aA{$\alpha$-nucleus\ }
\def\aap{($\alpha,\alpha'$)\ }

\def\aPb{$\alpha+^{208}$Pb\ }
\def\bbox#1{\mbox{\boldmath $#1$}}
\begin{document}
\begin{frontmatter}
\title{Microscopic study of the isoscalar giant resonances in $^{208}$Pb
induced by inelastic $\alpha$ scattering}
\author{Do Cong Cuong$^1$, Dao T. Khoa$^1$ and Gianluca Col\`o$^2$}
\address{$^{1}$Institute for Nuclear Science {\rm \&} Technique, VAEC \\
179 Hoang Quoc Viet Rd., Nghia Do, Hanoi, Vietnam. \\
$^2$ Dipartimento di Fisica, Universit\`a degli Studi di Milano,
and INFN, Sez. di Milano
\\ via Celoria 16, 20133, Milano, Italy.}

\begin{abstract}
The energetic beam of (spin and isospin zero) $\alpha$-particles remains a very
efficient probe for the nuclear isoscalar giant resonances. In the present work,
a microscopic folding model study of the isoscalar giant resonances in
$^{208}$Pb induced by inelastic \aPb scattering at $E_{\rm lab}=240$ and 386 MeV
has been performed using the (complex) CDM3Y6 interaction and nuclear transition
densities given by both the collective model and Random Phase Approximation
(RPA) approach. The fractions of energy weighted sum rule around the main peaks
of the isoscalar monopole, dipole and quadrupole giant resonances were probed in
the Distorted Wave Born Approximation analysis of inelastic \aPb scattering
using the double-folded form factors given by different choices of the nuclear
transition densities. The energy distribution of the $E0,\ E1$ and $E2$
strengths given by the multipole decomposition \mbox{analyses} of the \aap data
under study are compared with those predicted by the RPA calculation.
\end{abstract}

\end{frontmatter}

\section{Introduction}
\label{intro} Isoscalar giant resonances \cite{books} in medium and heavy nuclei
are the pronounced manifestation of nuclear collective motion and, hence, they
carry important information about the dynamics of the nuclear excitation process
and the properties of the nuclear Hamiltonian. Although their systematic study
started more than three decades ago, a number of challenging questions still
remain open. For example, the observed compressional modes, like the $L=0$
isoscalar giant monopole resonance (ISGMR) or the $L=1$ isoscalar giant dipole
resonance (ISGDR), provide the optimal route to determine the nuclear matter
incompressibility $K_\infty$, a key quantity specifying the equation of state of
nuclear matter. However, the compatibility of the $K_\infty$ values deduced from
these two collective compressional modes is still under discussion (see, e.g.,
Refs.~\cite{Shlo06,Colo08} and references therein). The isoscalar giant
quadrupole resonance (ISGQR) and the isovector giant dipole resonance (IVGDR)
are well known in stable nuclei but it is remains unclear how these two modes
evolve in the unstable neutron-rich nuclei.

Since the first observation of ISGMR in $^{208}$Pb in the 70's of the last
century \cite{Har77,Har79,You77}, this compressional excitation mode of
$^{208}$Pb has been investigated in numerous experimental and theoretical
studies due to its fundamental importance in the determination of the nuclear
matter incompressibility $K_\infty$. Among the experimental tools, the spin and
isospin zero $\alpha$-particle remains the best probe for the isoscalar giant
resonances with $\Delta S=\Delta T=0$ and the most pronounced observations of
the ISGMR in $^{208}$Pb have been made so far in the \aap experiments, like the
recent high-precision measurements of inelastic \aPb scattering at $E_{\rm
lab}=240$ MeV by the Texas A\&M University group \cite{You04} and 386 MeV by the
Osaka group \cite{Uch03,Uch04}. The exploratory theoretical studies of the
isoscalar dipole mode were done in the early 80s by M.N. Harakeh and A.E.L.
Dieperink \cite{Dieperink}, as well as by N. Van Giai and H. Sagawa \cite{giai},
but the first direct observation of the ISGDR in the
$^{208}$Pb$(\alpha,\alpha')$ reaction has been made only years later by Davis
\emph{et al.} \cite{Dav97}. Like the ISGMR, the knowledge about the ISGDR is of
vital importance for the determination of the nuclear incompressibility
$K_\infty$ \cite{Shlo06,Colo08}. Therefore, both the \aap experiments at 240 MeV
\cite{You04} and 386 MeV \cite{Uch03,Uch04} were aimed at an accurate
measurement of the ISGDR strength distribution in $^{208}$Pb. One of the main
problems in the experimental study of giant resonances has been, and still is,
the difficulty to disentangle different modes when their energies overlap. For
example, the excitation energy of ISGMR in $^{208}$Pb has been accurately
determined from the most forward part of the \aap cross section measured at 240
MeV to be $E_0\approx 13.96\pm 0.20$ MeV \cite{You04} while the \aap experiment
at 386 MeV has deduced $E_0\approx 13.4\pm 0.2$ MeV \cite{Uch04} using
essentially the same method of multipole decomposition analysis (MDA) to
disentangle the ISGMR peak from a mixed spectrum of different ($\Delta S=\Delta
T=0$) excitations including, in particular, the low-energy peak of ISGDR at an
excitation energy around 13 MeV \cite{You04,Uch04}.

Moreover, there exists a basic problem when experimental results need to be
compared with those predicted by a theoretical structure model. It is common, in
the theoretical works, to calculate the strength function $S(E)$ associated with
a given nuclear transition operator $Q$, that is,
\begin{equation}
 S(E)=\sum_\nu \delta(E-E_\nu)\ | < \nu | Q | 0 > |^2,
\label{st1}
\end{equation}
where $\nu$ labels a complete set of final states which can be excited by acting
$Q$ upon the ground state $| 0 >$. In terms of single-particle degrees of
freedom, the isoscalar $L$-multipole operator is given by
\begin{equation}
 Q_{LM} = \sum_{i=1}^A r_i^L Y_{LM}(\hat{r}_i). \label{qlm}
\end{equation}
In the case of the isoscalar monopole and dipole modes one has to replace
$r_i^L$ by $r_i^{L+2}$. The predicted strength function (\ref{st1}) is then used
to compare with the corresponding experimental strength distribution deduced for
the considered $L$-multipole isoscalar excitation. In reality, however, the
measured inelastic scattering data are inclusive spectra over a wide energy
range which contain the strengths of isoscalar excitations with different
multipolarities as well as the contamination by the continuum background. The
`experimental' strength function for a given $L$-multipole excitation depends
strongly, therefore, on both the method to exclude continuum background and the
MDA to disentangle contribution of a given multipole from the inelastic
scattering spectrum. As a result, the comparison between the theoretical and
experimental strength functions can be made in many cases only qualitatively.

Nevertheless, a more direct and quantitative comparison between theory and
experiment is possible if the nuclear structure information is accurately
included into a microscopic description of the inelastic \aap scattering cross
section measured for a given peak of the resonance energy, within either the
distorted-wave Born approximation (DWBA) or coupled-channel formalism
\cite{Sat83}. The well-known portal of this procedure is the double-folding
model (DFM) which uses the nuclear ground state and transition densities of the
$\alpha$-projectile and target nucleus and an effective nucleon-nucleon ($NN$)
interaction to calculate the \aA optical potential (OP) and inelastic scattering
form factors (FF) of different multipolarities for the DWBA calculation (see,
e.g., Ref.~\cite{Kho00} and references therein). We note that the single-folding
method \cite{KhoSat97} has been widely used to calculate the \aA inelastic
scattering FF, using an appropriate $\alpha$-nucleon interaction and nuclear
transition densities given by the collective model, for the multipole
decomposition analysis of the experimental spectrum within DWBA
\cite{You04,Uch03,Uch04}. In this respect, the DFM calculation of the \aA
inelastic scattering FF using a realistic effective $NN$ interaction can be used
to probe the transition strength extracted from the MDA for a given isoscalar
excitation as well as the reliability of different choices for the nuclear
transition densities. The latter aspect is quite essential because the
collective model transition density has been shown to be reasonable for the
collective modes which are, as a rule, concentrated in a limited energy region
while other parts of the spectrum could be dominated by pure particle-hole
states or states with intermediate character.

We note further that the existing structure models have been substantially
improved in recent years and the nuclear linear response theory has been used
successfully to describe the excitation of vibrational modes. In a magic nucleus
like $^{208}$Pb where the pairing does not manifest itself, this linear response
approach is also known as Random Phase Approximation (RPA) method. RPA has been
formulated many years ago, but only recently the fully selfconsistent
calculations without crude approximations have become available. Therefore, it
is also timely to study the inelastic \aA scattering data measured recently for
the known isoscalar modes to see how the RPA nuclear wave functions can be
probed in the folding + DWBA analysis of these data. To this goal, we have
performed in the present work a detailed folding model analysis of the
high-precision inelastic \aPb scattering data measured at $E_{\rm lab}=240$ MeV
\cite{You04} and 386 MeV \cite{Uch03,Uch04}, using a \emph{complex} version of
the density dependent CDM3Y6 interaction \cite{Kho97} and nuclear transition
densities given by both the collective model (CM) and the RPA calculation. After
a brief overview of the theoretical formalism in Sec.~\ref{forma}, results of
the DFM + DWBA analysis of the considered \aap data using the CM transition
densities are presented and discussed in Sec.~\ref{sec3}. The RPA description of
the $E0,\ E1$ and $E2$ strength distributions and DFM + DWBA results given by
the RPA transition densities are discussed in Sec.~\ref{sec4}. The main
conclusions and perspectives are given in the Summary.

\section{Formalism}
\label{forma}

In this Section, we briefly describe the theoretical model used to calculate the
inelastic \aap cross sections in the DWBA. As mentioned above, our microscopic
study of the inelastic \aPb scattering is based on the double-folding model
\cite{Kho00} which uses the nuclear ground state and transition densities of the
$\alpha$-projectile and target nucleus and an appropriate effective $NN$
interaction to calculate the \aA OP and inelastic scattering FF for the DWBA
analysis. The nuclear structure information on the isoscalar giant resonances in
$^{208}$Pb is embedded in the nuclear transition densities used in the folding
calculation of the inelastic scattering FF. Two choices of the nuclear
transition densities were used in the present work: the phenomenological
transition densities given by the collective model (discussed below in
Sec.~\ref{coll}) and microscopic transition densities given by the RPA approach
(discussed in Sec.~\ref{RPA}).

\subsection{Effective density dependent $NN$ interaction}
\label{inter}

Among various choices of the effective $NN$ interaction, a density dependent
version of the M3Y interaction (dubbed as CDM3Y6 interaction \cite{Kho97}) has
been used successfully in the folding model analyses of the (refractive) elastic
and inelastic \aA scattering (see the recent review in Ref.~\cite{Kho07}). The
density dependent parameters of the CDM3Y6 interaction were carefully adjusted
in the Hartree-Fock (HF) scheme to reproduce the saturation properties of
nuclear matter \cite{Kho97}. The first version of the CDM3Y6 interaction is
\emph{real} and can be used to predict the real OP and inelastic scattering FF
only. To avoid a phenomenological choice of the imaginary OP and inelastic
scattering FF, we have supplemented the real CDM3Y6 interaction with a realistic
\emph{imaginary} density dependence whose parameters were determined based on
the Brueckner Hartree-Fock (BHF) results for the nucleon OP in nuclear matter by
Jeukenne, Lejeune and Mahaux (JLM) \cite{Je77}. It has been shown in our recent
work \cite{Kho08} that the same form of the CDM3Y functional \cite{Kho97} can be
used to obtain the density dependence of the \emph{imaginary} term. Thus, the
complex CDM3Y6 interaction used in the present folding model analysis is
determined as
\begin{eqnarray}
{\rm Re (Im)}\ v_{\rm D(EX)}(E,\rho,s)=F_{\rm R(I)}(E,\rho)v_{\rm D(EX)}(s),
\label{g1} \\
F_x(E,\rho)=C_x[1+\alpha_x \exp(-\beta_x\rho)-\gamma_x\rho],\ x={\rm R,I}.
\label{g2}
\end{eqnarray}
The radial parts of the direct and exchange interactions $v_{\rm D(EX)}(s)$ were
kept unchanged, as derived \cite{Kho00} from the M3Y interaction based on the
G-matrix elements of the Paris $NN$ interaction \cite{An83}, in terms of three
Yukawas
\begin{eqnarray}
 v_{\rm D}(s) &=& 11061.625{{\exp(-4s)}\over{4s}}-
 2537.5{{\exp(-2.5s)}\over{2.5s}}, \\
 v_{\rm EX}(s) &=& -1524.25{{\exp(-4s)}\over{4s}}-518.75{{\exp(-2.5s)}\over{2.5s}}
 -7.8474{{\exp(-0.7072s)}\over{0.7072s}}. \nonumber
\label{g3}
\end{eqnarray}
While parameters of the real density dependence $F_{\rm R}$ were taken from the
original HF calculation of nuclear matter \cite{Kho97}, those of the imaginary
density dependence $F_{\rm I}$ were adjusted iteratively until the HF result for
the imaginary nucleon OP in nuclear matter agrees reasonably with the JLM result
\cite{Je77} as well as the shape of imaginary folded OP becomes close to the
phenomenological Woods-Saxon imaginary OP found at each energy. All parameters
of the complex density dependence are given in Table~1. We note that the dynamic
change in the density dependence $F_x(\rho)$ caused by the excitation of the
target is taken into account properly in the folding calculation using method
given in Ref.~\cite{Kho00}.

\subsection{Double-folding model}
\label{folding}

The generalized double-folding model of Ref.~\cite{Kho00} was used to evaluate
the complex \aA OP and inelastic scattering FF from the following HF-type matrix
elements of the CDM3Y6 interaction (\ref{g1})-(\ref{g2}) between the projectile
nucleon $i$ and target nucleon $j$
\begin{equation}
 U_{A\to A^*}=\sum_{i\in \alpha;j\in A,j'\in A^*}[<ij'|v_{\rm D}|ij>
 + <ij'|v_{\rm EX}|ji>],
\label{fd1}
\end{equation}
where $A$ and $A^*$ are states of the target in the entrance- and exit channel
of the \aA scattering, respectively. Thus, Eq.~(\ref{fd1}) gives the (diagonal)
elastic OP if $A^*=A$ and (nondiagonal) inelastic scattering FF if otherwise.
The (local) direct term is readily evaluated by the standard double-folding
integration
\begin{eqnarray}
 U_{\rm D}(E,\bbox{R})=\int\rho_\alpha(\bbox{r}_\alpha)
 \rho_A(\bbox{r}_A)v_{\rm D}(E,\rho,s)d^3r_\alpha d^3r_A,\nonumber \\
 \bbox{s}=\bbox{r}_A-\bbox{r}_\alpha+\bbox{R}.
\label{fd2}
\end{eqnarray}
The antisymmetrization gives rise to the exchange term in Eq.~(\ref{fd1}) which
is, in general, nonlocal in the coordinate space. However, it has been shown
\cite{Kho00,Kho07} that an accurate local equivalent exchange potential can be
obtained using the local WKB approximation \cite{Sat83} for the change in
relative motion induced by the exchange of spatial coordinates of each
interacting nucleon pair
\begin{eqnarray}
U_{\rm EX}(E,\bbox{R}) =\int \rho_\alpha
(\bbox{r}_\alpha,\bbox{r}_\alpha+\bbox{s})\rho_A(\bbox{r}_A,\bbox{r}_A
-\bbox{s})v_{\rm EX}(E,\rho,s) \nonumber \\
\times\exp\left({i\bbox{K}(\bbox{R})\bbox{s}}\over{M}\right)d^3r_\alpha d^3r_A.
\label{fd3}
\end{eqnarray}
Here $\bbox{K}(\bbox{R})$ is the local momentum of relative motion determined
from
\begin{equation}
 K^2(\bbox{R})={{2\mu}\over{\hbar}^2}[E_{\rm c.m.}-{\rm Re}~U_0(E,\bbox{R})-
 V_C(\bbox{R})],
\label{fd4}
\end{equation}
where $\mu$ is the reduced mass, $M=4A/(4+A)$, $E_{\rm c.m.}$ is the scattering
energy in the center-of-mass (c.m.) frame, $U_0(E,\bbox{R})$ and $V_C(\bbox{R})$
are the nuclear and Coulomb parts of the \emph{real} \aA OP, respectively. The
calculation of $U_{\rm D(EX)}$ is done iteratively based on a density-matrix
expansion method \cite{Kho00,Kho01}. All technical details of the folding
calculation of $U_{\rm D(EX)}$ are the same as those given in Ref.~\cite{Kho00},
excepting the use of a realistic local approximation for the \emph{transition}
density matrix suggested by Love \cite{Lo78} and a recoil correction to the
exchange term (\ref{fd3}) suggested by Carstoiu and Lassaut \cite{Car96}.

\subsection{Collective model for the nuclear transition densities}
\label{coll}

To calculate consistently both the OP and inelastic scattering FF for the \aPb
system one needs to represent the target density in terms of the ground state
(g.s.) and transition parts as $\rho(\bbox{r})=\rho_0(r)+\delta\rho(\bbox{r})$.
The explicit expression of the inelastic scattering FF for a given isoscalar
excitation (see Ref.~ \cite{Kho00}) can be deduced from the double-folding
integrals (\ref{fd2})-(\ref{fd3}), using the following multipole decomposition
of $\delta\rho(\bbox{r})$
\begin {equation}
 \delta\rho(\bbox{r})=\sum_{LM}C_L\delta\rho_L(r)[i^L Y_{LM}(\hat{\bbox{r}})]^*,
 \label{fd5}
\end{equation}
where $C_0=\sqrt{4\pi}$ and $C_L$=1 for $L\neq 0$. Given the strong collective
nature of the isoscalar giant resonances, macroscopic methods to construct the
nuclear transition density of the $2^L$-pole isoscalar excitation
$\delta\rho_L(r)$, based on the collective model, are widely used in the folding
model calculation \cite{KhoSat97,Hor95} and multipole decomposition analysis
\cite{Har79,You77,You04,Uch03,Uch04} of the \aap data.

For the isoscalar giant resonances with $L\geq 2$, we adopt the so-called
Bohr-Mottelson prescription \cite{Bohr} to construct the transition densities
\begin {equation}
\delta\rho_L(r)=-\delta_L\frac{d\rho_0(r)}{dr}. \label{fd6}
\end{equation}
Here $\rho_0(r)$ is the g.s. density and $\delta_L$ is the deformation length of
the considered isoscalar excitation. The g.s. density of $^{208}$Pb was taken as
a Fermi distribution with parameters \cite{Far85} chosen to reproduce the
shell-model density for $^{208}$Pb. Within the \emph{isoscalar} assumption
\cite{Kho00,Hor95} the same deformation length $\delta_L$ is employed, as a
rule, for both the neutron and proton parts of the nuclear transition density
(\ref{fd6}). For the low-lying excitations, like the first $3^-$ state in
$^{208}$Pb considered below, the deformation length is normally determined
\cite{Kho00} from the measured electric transition strength $B(EL)$. In terms of
the energy weighted sum rule (EWSR) for the operator (\ref{qlm}), if a single
state $|\nu>$ at the excitation energy $E_\nu$ exhausts 100\% of the isoscalar
EWSR then the corresponding deformation length is determined \cite{Hor95} as
\begin{equation}
 \delta_L^2(E_\nu)= {\hbar^2\over 2m} \frac{4\pi}{AE_\nu}\frac{L(2L+1)^2}
{(L+2)^2}\frac{<r^{2L-2}>}{<r^{L-1}>^2};\
<r^{L-1}>=\frac{\int\rho_0(r)r^{L+1}dr}{\int\rho_0(r)r^2dr}.
%\nonumber
 \label{fd7}
\end{equation}
In the case of ISGMR, the pure breathing mode (or scaling) assumption
\cite{Uberall} is used to construct the nuclear transition density
\begin{equation}
\delta\rho_0(r) = -\delta_0\left[3\rho_0(r)+r\frac{d\rho_0(r)}{dr} \right].
\label{fd8}
\end{equation}
If an isoscalar monopole state $|\nu>$ at the excitation energy $E_\nu$ exhausts
100\% of the monopole EWSR then its deformation length is determined
\cite{Hor95} as
\begin{equation}
\delta_0^2(E_\nu)= \frac{\hbar^2}{2m}\frac{4\pi}{AE_\nu}\frac{1}{<r^2>}.
\label{fd9}
\end{equation}

Another special case is that of the isoscalar dipole excitation for which a
macroscopic model based on the compressional hypothesis, with a proper
 center-of-mass subtraction, has been suggested by Harakeh and Dieperink
\cite{Dieperink}. Dropping the high-order term $\epsilon$ which is negligible
for $A\geq 20$ \cite{Dieperink}, the transition density of an isoscalar dipole
state is written as
\begin{equation}
\delta\rho_1(r)=-\frac{\delta_{1}}{R}\left[3r^2{d\over dr}+ 10r-{5\over 3}<r^2>
{d\over dr}\right]\rho_0(r), \label{fd10}
\end{equation}
where $R$ is the half-density radius of the g.s. density distribution
$\rho_0(r)$. If an isoscalar dipole state $|\nu>$ at the excitation energy
$E_\nu$ exhausts 100\% of the EWSR for the dipole operator, with spurious c.m.
oscillation subtracted \cite{Dieperink,giai}, then its deformation length is
\begin{equation}
 \delta_{1}^2(E_\nu)=\frac{6\pi\hbar^2}{mAE_\nu} R^2 \left[11<r^4>-{25\over 3}
 <r^2>^2 \right]^{-1}. \label{fd11}
\end{equation}
The CM transition densities (\ref{fd6})-(\ref{fd11}) are normalized in our
calculation to describe the excitation process $|{\rm g.s.}>\to |\nu>$,
that is, they correspond to the upward transition amplitude. In this way, the
corresponding isoscalar transition strength is $S_L=|M_L|^2$ with the transition
moment determined as
\begin{eqnarray}
M_{L}=\int dr\ r^{L+2}\delta\rho_{L}(r)
 \ \ \ \ & {\rm if} & L\geq 2, \nonumber \\
M_{L}= \int dr\ r^{4}\delta\rho_{L}(r)
 \ \ \ \ & {\rm if} & L=0, \nonumber \\
M_{L} =\int dr\left(r^{3}-{5\over 3}<r^2> r\right)r^2
 \delta\rho_{L}(r) \ \ \ \ & {\rm if}& L=1. \label{fd12}
\end{eqnarray}

\subsection{Microscopic RPA transition densities}
\label{RPA}

Despite a certain success of the collective model transition densities in
numerous folding model studies of the isoscalar giant resonances induced by
inelastic \aA scattering, there is no firm experimental evidence validating
their use. In the case of ISGMR, for example, there are only some results of
structure calculations showing that (\ref{fd8}) is a good representation of the
ISGMR transition density in the surface region \cite{Woude99}. Moreover, the
radial shapes of the collective model transition densities are assumed to be
independent of the excitation energies, which is surely not the case in the
reality. Therefore, the folding model analysis based on the transition densities
(for a given isoscalar mode) calculated selfconsistently at different excitation
energies by a microscopic RPA or quasiparticle RPA (QRPA) approach is expected
to provide a complementary and useful insight. While the QRPA transition
densities have been shown to give reasonably good results in many cases like,
e.g., in the folding model study of the lowest 2$^+$ states in the Sulfur
isotopes induced by inelastic proton scattering \cite{khoapp}, QRPA or RPA do
not systematically provide good results for the low-lying isoscalar excitations
when strong anharmonic effects are present. In this sense, RPA is expected to be
more suitable for giant resonances and it is, therefore, of interest to probe
the RPA transition densities in the present study. We note in this context a
similar attempt done recently to study the charge exchange ($^3$He,$t$) reaction
\cite{Guillot}. In general, the full coupling between the microscopic structure
and reaction models should be, in our opinion, pursued more extensively.

In the present calculations, we have chosen the parametrization set SLy5
\cite{Chabanat} of the Skyrme interaction for the RPA calculation of the
isoscalar states in $^{208}$Pb. We first solve the HF equations in the
coordinate space to construct the single-particle basis. All the radial
integrals are computed up to a maximum radius of 22.5 fm, using a mesh of 0.15
fm. The unoccupied single-particle states, including those at positive energies,
are obtained by putting the system in a large box of 22.5 fm, that is, the
continuum is discretized. A basis of particle-hole ($ph$) configurations is then
built upon all occupied states, as well as the lowest unoccupied states with
increasing values of the principal quantum number $n$, for each allowed value of
($l,j$). The RPA matrix equations are then solved in this basis, which has been
checked to be large enough to ensure that the appropriate sum rules are
satisfied. The procedure has already been explained in Ref.~\cite{comex2}.

From the solutions of the RPA equations, the energies $E_\nu$ of the excited
states $|\nu>$ as well as their wave functions are readily obtained. The radial
transition density $\delta\rho_{L\nu} (r)$ associated with a given ($2^L$-pole)
RPA state $|\nu>$ is given by
\begin{equation}
 \delta\rho_{L\nu}^{(q)}(r) = \sum_{ph\in q}\left( X^{(L\nu)}_{ph} +
 Y^{(L\nu)}_{ph} \right) < p || Y_L || h > R_p(r)R_h(r), \label{rpa1}
\end{equation}
where $X$ and $Y$ are the forward and backward RPA amplitudes and $R(r)$ labels
the radial part of the single-particle wave function. The proton and neutron
parts (labelled by $q=p,n$) of the transition density (\ref{rpa1}) are computed
separately. The isoscalar transition strength of the RPA state $|\nu>$ is
evaluated as $S_{L\nu}=|M_{L\nu}|^2$, where the transition moment $M_{L\nu}$ is
\begin{eqnarray}
M_{L\nu}=\int dr\ r^{L+2}\left[\delta\rho_{L\nu}^{(p)}(r)+
\delta\rho_{L\nu}^{(n)}(r)\right] \ \ \ \ & {\rm if} & L\geq 2,
\nonumber \\
M_{L\nu}= \int dr\ r^{4} \left[\delta\rho_{L\nu}^{(p)}(r)+
\delta\rho_{L\nu}^{(n)}(r) \right]\ \ \ \ & {\rm if} & L=0,
\nonumber \\
M_{L\nu} =\int dr\left(r^{3}-{5\over 3}<r^2> r\right)r^2
 \left[\delta\rho_{L\nu}^{(p)}(r)+\delta\rho_{L\nu}^{(n)}(r)\right]
 \ \ \ \ & {\rm if}& L=1. \label{rpa2}
\end{eqnarray}

\section{Double-folding model + DWBA analysis using the collective model
 transition densities}
\label{sec3}

\begin{figure}[bht]
 \begin{center}\vspace{-2cm}\hspace{-1cm}
\includegraphics[angle=0,scale=0.70]{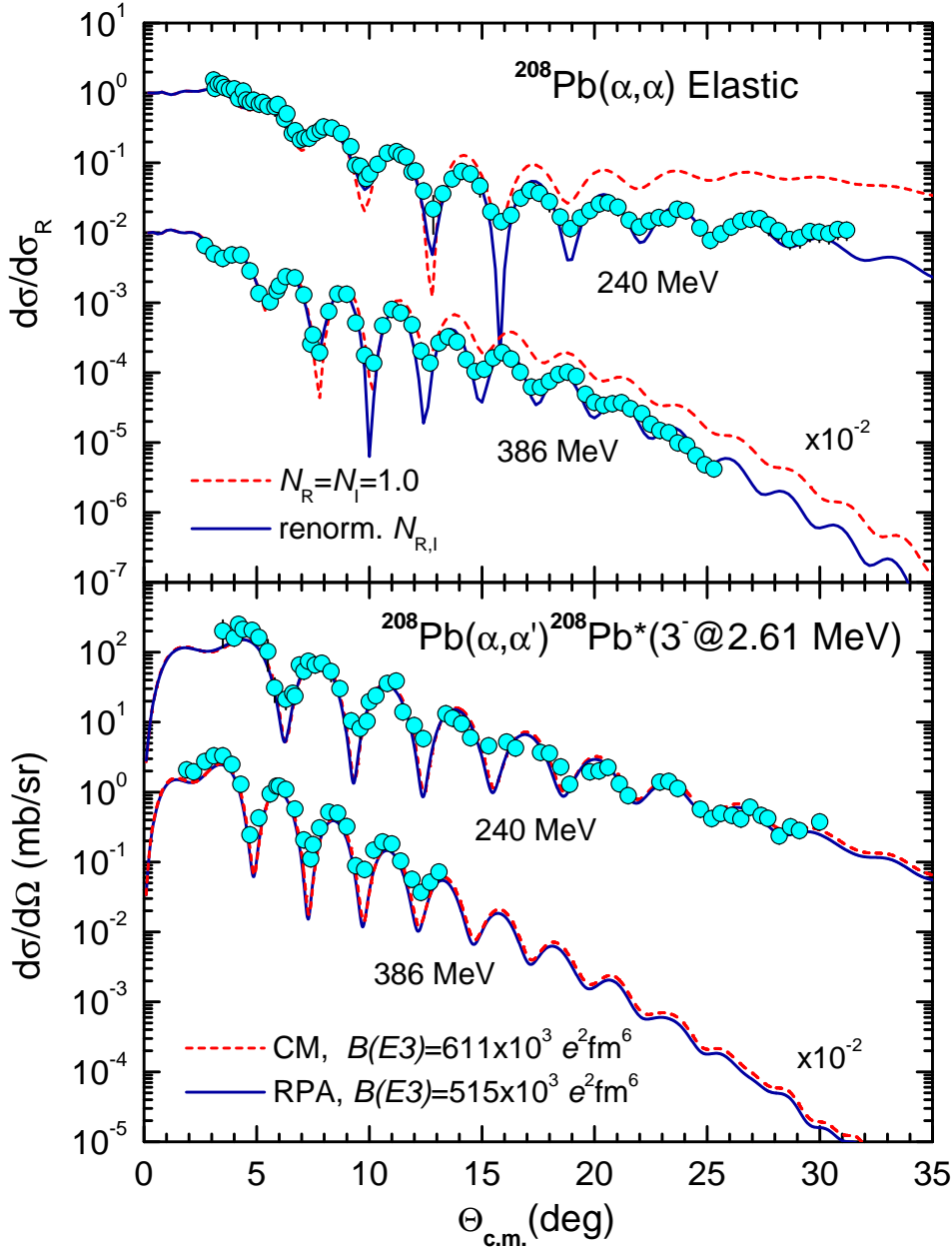}\vspace{-1cm}
\caption{Upper panel: OM description of the elastic \aPb scattering data at 240
MeV \cite{You04} and 386 MeV \cite{Uch03,Uch04} given by the unrenormalized
(dash curves) and renormalized complex folded OP (solid curves). Lower panel:
DWBA descriptions of the inelastic \aPb scattering data \cite{You04,Uch03,Uch04}
for $3^-_1$ state of $^{208}$Pb ($E_x = 2.61$ MeV) given by the collective model
(dash curves) and RPA (solid curves) nuclear transition densities.}\label{f1}
\end{center}
\end{figure}

To generate realistic distorted waves for the DFM + DWBA study of the isoscalar
giant resonances, we first used the nuclear g.s. densities of $^4$He and
$^{208}$Pb taken from Refs.~\cite{Sa79} and \cite{Far85}, respectively, to
calculate the complex folded OP for the optical model (OM) analysis of the
elastic \aPb scattering data at $E_{\rm lab}=240$ MeV \cite{You04} and 386 MeV
\cite{Uch03}. To fine tune the complex strength of the CDM3Y6 interaction
(\ref{g1}), renormalization coefficients $N_{\rm R}$ and $N_{\rm I}$ of the real
and imaginary elastic folded potentials (\ref{fd1}) were adjusted by the OM fit
to the elastic data at each energy (see OM results shown in upper panel of
Fig.~\ref{f1}). One can see from Table~\ref{t1} that the best-fit $N_{\rm R}$
coefficient is rather close to unity. The best-fit $N_{\rm I}$ of about 1.4 is
reasonable because the imaginary strength of the CDM3Y6 interaction was tuned to
the BHF results for nuclear matter and gives, therefore, only the ``volume"
absorption. To effectively account for the surface absorption caused by
inelastic scattering and transfer reactions, an enhanced $N_{\rm I}$ coefficient
is naturally needed (compare the dash and solid curves in upper part of
Fig.~\ref{f1}). Our OM calculation also predicted the total reaction cross
sections $\sigma_{\rm R}$ very close to the experimental values measured at the
nearby energies. Thus, the elastic distorted waves given by the present DFM
calculation should be accurate for the DWBA analysis of inelastic \aPb
scattering.

For the inelastic scattering form factor, a standard method used so far in the
DFM + DWBA analyses of inelastic \aA scattering \cite{Kho00,KhoSat97} is to
scale the real and imaginary inelastic folded FF by the same renormalization
coefficients $N_{\rm R}$ and $N_{\rm I}$ as those deduced from OM analysis of
elastic scattering data. We show in lower panel of Fig.~\ref{f1}, as
illustration, the DWBA description of inelastic \aPb scattering data for $3^-_1$
state of $^{208}$Pb given by the inelastic folded FF scaled by the same
coefficients $N_{\rm R}$ and $N_{\rm I}$ as those given in Table~\ref{t1}. By
using a deformation length $\delta_L$ of the CM transition density (\ref{fd6})
chosen to reproduce the measured transition rate $B_{\rm exp}(E3)\approx
611\times 10^3\ e^2$ fm$^6$ \cite{Kib02}, a very satisfactory description of the
inelastic scattering data for $3^-_1$ state of $^{208}$Pb has been obtained.
About the same good DWBA description was also obtained with the microscopic
nuclear transition density (\ref{rpa1}) given by the RPA calculation, without
any {\em ad hoc} adjustment. We note that the Coulomb part of inelastic
scattering FF is obtained in the present work by double folding the proton parts
of the $^4$He g.s. density and $^{208}$Pb transition density with the Coulomb
interaction, using a folding method similar to that used for the nuclear part.

\subsection{MDA and deformation lengths for the CM transition densities}
 \label{sec3.0}
Before discussing the DFM + DWBA results for the isoscalar giant resonances we
briefly recall here how the experimental transition strengths are determined
from the multipole decomposition analysis of the measured $(\alpha,\alpha')$
angular distributions \cite{You04,Uch03,Uch04}. At a given energy bin, the
measured (double) differential cross section is expressed within the MDA as a
superposition of the angular distributions calculated for different transferred
angular momenta $L$ as
\begin{equation}
\left[\frac{d^{2}\sigma}{d\Omega dE}(\Theta_{\rm c.m.},E_x)\right]^{\rm exp.} =
\sum_{L=0}^{L_{\rm max}}a_L(E_x)\left[\frac{d^{2}\sigma}{d\Omega dE}(\Theta_{\rm
c.m.},E_x) \right]^{\rm calc.}_L. \label{MDA}
\end{equation}
Here $[d^{2}\sigma/d\Omega dE]^{\rm calc.}_{L}$ is calculated within the DWBA
using the inelastic scattering FF generated from the appropriate CM nuclear
transition density (\ref{fd6})-(\ref{fd11}) by a single-folding method
\cite{KhoSat97}. The CM nuclear transition densities entering the MDA are first
determined with 100\% exhaustion of the corresponding EWSR (see
Sec.~\ref{coll}), then a least-$\chi^2$-fit procedure determines all $a_L(E_x)$
coefficients for the considered experimental energy bin. As a result, each
best-fit $a_L(E_x)$ coefficient represents the fraction of EWSR exhausted by the
corresponding isoscalar $2^L$-pole excitation mode in the energy bin under the
MDA analysis. In terms of deformation length $\delta_L$ for a given excitation
mode in the considered energy bin $a_L(E_x)=(\delta_L/\delta^{\rm max}_L)^2$,
where $\delta^{\rm max}_L$ is the maximum deformation length determined [see
Eqs.~(\ref{fd7}), (\ref{fd9}) and (\ref{fd11})] to exhaust 100\% of the
corresponding EWSR.

The MDA analysis of the inelastic \aPb scattering data at $E_{\rm lab}=240$ MeV
measured by the Texas A\&M University group \cite{You04} was done in the energy
bins of 640 or 800 keV width to deduce the isoscalar $EL$ strength distributions
over a wide range of excitation energy. The MDA of the 240 MeV data shows, in
particular, that the full exhaustion (around 100\%) of the isoscalar EWSR has
been observed for the ISGMR and ISGQR. The main ISGMR peak has been accurately
determined from the 240 MeV data to be at $E_x\approx 13.96\pm 0.20$ MeV with a
width $\Gamma\approx 2.88\pm 0.20$ MeV, and fragmentation of the $E0$ strength
up to about 20 MeV has been observed. The MDA analysis of high-precision
inelastic \aPb scattering data measured by the Osaka group at $E_{\rm lab}=386$
MeV \cite{Uch03,Uch04} (done in the energy bins of 1 MeV width) has shown a much
stronger fragmentation of the $E0$ strength over excitation energies well
above 30 MeV, and a less pronounced ISGMR peak (observed at $E_x\approx 13.4\pm
0.2$ MeV with a wider width $\Gamma\approx 4.0\pm 0.4$ MeV).

One could reproduce the ISGMR peaks observed in the \aap experiment at 240 MeV
\cite{You04} and 386 MeV \cite{Uch03,Uch04} in the microscopic structure models,
using either nonrelativistic or relativistic functionals which give
$K_\infty\approx 240\pm 20$ MeV \cite{Col04,Vre03}. This is why some consensus
has been reached \cite{Shlo06,Colo08} on this empirical value for $K_\infty$,
where the error of $\pm$ 20 MeV is not simply associated with the experimental
uncertainty on the ISGMR energy, but rather with our still incomplete
understanding of the structure of energy functionals (in particular, of their
density dependence). In this connection, we note that a pure {\em experimental}
discrepancy of 500 keV in the observed ISGMR peaks could also result in a
difference of $\Delta K_\infty\approx 20$ MeV in the $K_\infty$ values extracted
from empirical formulas relating the ISGMR peak in $^{208}$Pb and nuclear matter
incompressibility $K_\infty$ \cite{Colo08,Str82,Bla95}. Such a difference is
quite sizeable and hinders any further theoretical modelling of the energy
functionals.

If one uses a microscopic structure approach to determine $K_\infty$ from the
ISGMR data, the location of $E_{\rm ISGMR}$ will affect the deduced $K_\infty$
value. In about the same way, the observed ISGDR peak may also be directly
related to the $K_\infty$ value \cite{books,Shlo06,Colo08}. Given such a vital
importance of the ISGMR and ISGDR excitations in determining the nuclear matter
incompressibility $K_\infty$ and the fact that a simple single-folding method
\cite{KhoSat97} was used to calculate the \aPb inelastic scattering FF for the
MDA analyses of Refs.~\cite{You04,Uch03,Uch04}, we deem it necessary to probe
the ISGMR and ISGDR strength distributions extracted from these two experiments
again in our DFM + DWBA approach.

It is complementary to note that the density dependent CDM3Y6 interaction used
in the present DFM calculation was parametrized in the HF scheme to reproduce
$K_\infty=252$ MeV at the saturation density of symmetric nuclear matter
\cite{Kho97,Kho07}, and it has been successfully used in numerous OM and DWBA
analyses of elastic and inelastic \aA scattering. The isoscalar $EL$ strengths
(in terms of exhausted fractions of the corresponding EWSR for $L=0,1,2,3$)
given by the MDA analyses of Refs.~\cite{You04,Uch03,Uch04} for the main peaks
of the ISGMR, ISGQR and ISGDR together with those predicted by the RPA
calculation are presented in Table~\ref{t2}. Each energy bin in Table~\ref{t2}
has been chosen so that the strongest $EL$ strengths around the main resonance
peaks deduced from the two experiments can be used consistently in the same DFM
+ DWBA calculation. For example, the ISGMR peak was found by the MDA of 240 MeV
data \cite{You04} and 386 MeV data \cite{Uch03,Uch04} in the energy bins
centered at $E_x\approx 14.1$ and 13.5 MeV, respectively, and the corresponding
$E0$ strengths should be studied in the same DFM + DWBA analysis. The full
energy distributions of the $EL$ strengths for $L=0,1,2$ are shown below in
Sect.~\ref{sec4}.

\subsection{Isoscalar $EL$ strengths near the ISGMR peak} \label{sec3.1}

In Fig.~\ref{f2} the inelastic \aPb scattering cross section at $E_{\rm
lab}=240$ MeV measured for the 640 keV-wide energy bin centered at $E_x = 14.1$
MeV \cite{You04} are compared with the DFM + DWBA results given by the
collective model transition density (\ref{fd8})-(\ref{fd9}). In this energy bin,
the $E0$ strength deduced from the 240 MeV data is strongest and exhausts about
37.6\% of the $E0$ EWSR. The isoscalar $E1$ strength is quite significant (6.3\%
of the $E1$ EWSR) in this energy bin due to the dipole strength coming from the
low-energy peak of ISGDR located around 13 MeV and affecting significantly the
total angular distribution. The isoscalar $E2$ strength of about 6.6\% of the
$E2$ EWSR as well as no contribution from isoscalar $E3$ excitation were found
in this energy bin \cite{You04}. One can see in the lower panel of Fig.~\ref{f2}
that the measured inelastic scattering cross section is reasonably described by
the DFM + DWBA calculation using the CM transition densities scaled to the
isoscalar ($L=0,1,2$) strengths given by the MDA of Ref.~\cite{You04}. Given the
monopole and dipole angular distributions oscillating out of phase, a smooth
angular distribution seen in the 240 MeV data shows clearly the mixture of the
$E1$ strength from the low-energy peak of ISGDR in the considered energy bin.
Since the MDA of the Osaka data \cite{Uch03,Uch04} has given the strongest $E0$
strength in the 1 MeV energy bin centered at $E_x = 13.5$ MeV, we found it
appropriate to use the isoscalar $EL$ strengths deduced for this energy bin to
construct the CM transition densities for our DFM + DWBA analysis of the 240 MeV
data for $E_x = 14.1$ MeV. According to the MDA of the Osaka data, the $E0$
strength is strongly fragmented over a wide energy range and only about 16\% of
the $E0$ EWSR has been located in the energy bin around the ISGMR peak at $E_x =
13.5$ MeV. While the isoscalar dipole strength deduced from the 386 MeV data is
quite close to that deduced from the 240 MeV data, the isoscalar $E2$ strength
was found \cite{Uch03,Uch04} much stronger (up to 15\% of the $E2$ EWSR) in the
energy bin around $E_x = 13.5$ MeV. On top of that, about 3\% of the isoscalar
$E3$ EWSR strength was also observed by Uchida \emph{et al.} \cite{Uch03,Uch04}
in this energy bin. The DFM + DWBA description of the 240 MeV data for $E_x =
14.1$ MeV given by the CM transition densities scaled to the isoscalar
($0\leqslant L\leqslant 3$) strengths taken from Refs.~\cite{Uch03,Uch04} are
shown in upper panel of Fig.~\ref{f2}. Although a weaker $E0$ strength deduced
by Uchida \emph{et al.} \cite{Uch03,Uch04} gives a monopole cross section more
than 2 times smaller than that obtained with the $E0$ strength deduced by
Youngblood \emph{et al.} \cite{You04}, the overall description of the 240 MeV
data given by the isoscalar $EL$ transition strengths taken from
Refs.~\cite{Uch03,Uch04} remains satisfactory due to stronger $E2$ and $E3$
contributions. However, the lack of the $E0$ strength can still be seen in the
DWBA description of data points at the most forward angles shown in the upper
panel of Fig.~\ref{f2}.

It is natural also to check the DFM + DWBA description of the Osaka data at
$E_{\rm lab}=386$ MeV \cite{Uch03,Uch04,Uch03t} based on the same $EL$
transition strengths as discussed above. In Fig.~\ref{f3} the inelastic \aPb
scattering cross section at $E_{\rm lab}=386$ MeV measured for the 1 MeV-wide
energy bin centered at 13.5 MeV are compared with the DFM + DWBA results given
by the CM transition densities (\ref{fd8})-(\ref{fd9}) built upon the same
isoscalar $EL$ transition strengths as those used in Fig.~\ref{f2}. Except for
the two data points at the most forward angles which are fairly described by the
$EL$ strengths given by the MDA of the 240 MeV data \cite{You04}, the present
DFM + DWBA results strongly underestimate the measured data over the whole
angular range. About the same picture has been found for the energy bin centered
at $E_x = 14.5$ MeV when the DFM + DWBA results obtained with the $EL$ strengths
deduced for this bin are compared with the inelastic \aPb scattering data at
$E_{\rm lab}=386$ MeV \cite{Uch03t}. Such a big gap between the calculated and
measured cross sections seen in Fig.~\ref{f3} is not unexpected because the
contributions by the excitation modes of higher multipoles ($L>3$) are not taken
into account in our DFM + DWBA calculation. We recall that the authors of
Refs.~\cite{Uch03,Uch04,Uch03t} were able to measured the \aap energy spectrum
for the lead target without any contamination from the instrumental background
by using the high-resolution magnetic spectrometer Grand Raiden, and the MDA
analysis of the 386 MeV data has been done for all multipoles up to $L_{\rm
max}=14$ \cite{Uch03,Uch04,Uch03t}. Since the number of fitting parameters is
quite large in this case, it is not excluded that some continuum background
coming from other quasi-elastic processes like the pickup/breakup reactions has
been simply approximated by the high-multipole terms in the MDA series.
Moreover, any strong particle-hole $EL$ excitation with $9\leqslant L\leqslant
14$ in the energy region around 13-14 MeV would be unlikely from the structure
point of view. In contrast to the Osaka experiment, the MDA of the 240 MeV data
by the Texas A\&M University group was done only after a broad continuum
background (presumably caused by the high-multipole $EL$ excitations and
pickup/breakup reactions) has been substracted \cite{You04}. As a result, the
MDA of the 240 MeV data has been performed with less fitting parameters. The
fact that present DFM + DWBA results describe the 240 MeV data reasonably using
the $EL$ strengths deduced by the Texas A\&M University group \cite{You04} seems
to indicate that the single-folding method \cite{KhoSat97} used in their MDA of
inelastic \aA scattering is quite reliable. However, a closer inspection of the
lower part of Fig.~\ref{f2} and Fig.~5 in Ref.~\cite{You04} shows that the
present DFM + DWBA results are slightly underestimating the data points compared
to the MDA results obtained with the single-folding method. To explore such a
difference in the case of 386 MeV data, we have compared in Fig.~\ref{f4} the
present DFM + DWBA results with those given by the MDA of the \aap data measured
for the energy bin centered at $E_x = 14.5$ MeV \cite{Uch03,Uch04,Uch03t} and
found that the DFM + DWBA cross sections are indeed lower than those given by
the MDA (with the relative difference in the calculated total cross sections
ranging from  $\sim 57\%$ at the most forward angles to $\sim 44\%$ at
$\Theta_{\rm c.m.}\approx 14^\circ$). In particular, the difference between the
DFM + DWBA cross section for $L=2$ and that given by the MDA is very alarming,
as it is $\sim 185\%$ at the most forward angles and $\sim 69\%$ at $\Theta_{\rm
c.m.}\approx 14^\circ$ (see dash-dotted curves in Fig.~\ref{f4}). Such a
difference is quite significant and can result in sizable differences in the
$EL$ transition strengths deduced from the MDA of inelastic \aA scattering using
either single- or double-folding method. Since the present DFM approach is much
more advanced compared to the single-folding method \cite{KhoSat97} used so far
in the MDA, it is not excluded that the $EL$ strengths near the ISGMR or ISGQR
peaks deduced from the MDA are somewhat underestimated. In any case, the use of
the present DFM approach in the MDA of future \aA scattering data is strongly
recommended. We note further that while the ISGMR peaks observed in the two \aap
measurements seem to be in a reasonable agreement with the existing database
\cite{books,Shlo06,Colo08}, the difference in the observed distributions of the
isoscalar monopole strength is striking (about a factor of 2 in the most
important energy interval). It remains, therefore, an interesting challenge to
future experiments to confirm whether the $E0$ strength is mainly localized at
the excitation energies below 20 MeV \cite{You04} or widely fragmented to
energies beyond 30 MeV \cite{Uch03,Uch04}.

\subsection{Isoscalar $EL$ strengths near the ISGQR peak} \label{sec3.2}

The ISGQR at $E_x\approx 10\sim 11$ MeV  in $^{208}$Pb is perhaps one of the
most studied isoscalar giant resonances in nuclei. Nevertheless, like in the
case of the ISGMR, the isoscalar $E2$ strength distributions observed in the
\aap experiments at 240 MeV \cite{You04} and 386 MeV \cite{Uch03,Uch03t} are
sizeably different. For example, the $E2$ strength has been shown by the MDA of
the 240 MeV data to be concentrated mainly near the ISGQR peak and slightly
spread over the energies below 21 MeV, exhausting $100\pm 13$\% of the $E2$ EWSR
\cite{You04}. In contrast, the $E2$ strength given by the MDA of the 386 MeV
data \cite{Uch03,Uch03t} is broadly spread from about the same ISGQR peak to
energies beyond 30 MeV and, hence, exhausts more than 200\% of the $E2$ EWSR
(based on a direct integration of the tabulated $E2$ strength made available to
us by the authors).

In Fig.~\ref{f5} the inelastic \aPb scattering data at $E_{\rm lab}=240$ MeV
measured for the energy bin centered at $E_x = 10.3$ MeV \cite{You04} are
compared with the DFM + DWBA predictions based on the $EL$ strengths taken from
Refs.~\cite{You04,Uch03,Uch03t}. In this energy bin, the $E2$ strength deduced
from the 240 MeV data is stronger than that deduced from the 386 MeV data and
exhausts about 20\% of the $E2$ EWSR, and the DWBA description of the measured
angular distribution is better (see lower panel of Fig.~\ref{f5}) if the CM
transition densities are scaled to the $EL$ strength deduced from the 240 MeV
data \cite{You04}. We note that a weak isovector $E1$ strength was also included
into the MDA of the 240 MeV data to achieve a better DWBA fit to the data
measured for the two energy bins centered at 10.3 and 14.1 MeV \cite{You04}.
Given a dominant contribution by the isoscalar $EL$ strengths ($0\leqslant
L\leqslant 3$) to the \aap cross sections (see lower panels of Figs.~\ref{f2}
and \ref{f5}), we have chosen not to include the isovector $E1$ mixing into the
present DFM + DWBA calculation in order to show explicitly the role of the
isoscalar excitation modes. We have also observed (lower panel of Fig.~\ref{f5})
that the present DFM + DWBA results slightly underestimate the data points
compared to the MDA results of Ref.~\cite{You04} and this effect should be due
to the use of single-folding method (see further discussion below).

In the energy bin centered at $E_x = 10.9$ MeV for the 240 MeV data (or $E_x = 10.5$
MeV for the 386 MeV data) the $E2$ strengths deduced from both data sets are quite
close to each other, exhausting about 19 to 23\% of the $E2$ EWSR (see
Table~\ref{t2}). In addition, similar $E0$ and $E1$ strengths (from 3 to 4 \% of
the corresponding EWSR) were also deduced from both experiments. The only
difference is a significantly larger $E3$ strength deduced by the MDA of the 240
MeV data for this energy bin. As there is no \aap cross section measured at
$E_{\rm lab}=240$ MeV available to us for this energy bin, we compare in
Fig.~\ref{f6} the \aap data at $E_{\rm lab}=386$ MeV measured for the energy bin
centered at $E_x = 10.5$ MeV \cite{Uch03t} with the DFM + DWBA results given by
the isoscalar $EL$ strengths taken from Refs.~\cite{You04,Uch03,Uch04}. One can
see in lower panel of Fig.~\ref{f6} that the DFM + DWBA results based on the
$EL$ strengths taken from the MDA of the 240 MeV data \cite{You04} are much
closer to the data points compared with similar results for the \aap cross
sections at $E_{\rm lab}=386$ MeV shown in lower panels of Figs.~\ref{f3} and
\ref{f4}. This could well indicate a much weaker contribution by the
high multipoles to the \aap cross section at the excitation energy
around 10 MeV. Like the results obtained above for the excitation energies around 14
MeV, the DFM + DWBA calculation based on the $EL$ strengths taken from the MDA
of the 386 MeV data \cite{Uch03,Uch04} underestimates the data over the whole
angular range (see upper panels of Figs.~\ref{f3}, \ref{f4} and \ref{f6})
including the smallest angles where high multipoles are not expected to play a
major role.

\subsection{Isoscalar $EL$ strengths near the main ISGDR peak} \label{sec3.3}

The ISGDR in $^{208}$Pb has been observed in both the \aap experiments at 240
MeV \cite{You04} and 386 MeV \cite{Uch03,Uch04}. The isoscalar $E1$ strength
distribution has been shown clearly by these two experiments to split into two
parts: a weak low-energy peak centered at $E_x\approx 13$ MeV and the main,
broad high-energy peak at $E_x\approx 22.5$ MeV. In contrast to the ISGMR case,
the ISGDR peaks observed in these two measurements are quite close to each
other, except for some difference in the width deduced for the low-energy $E1$
peak. Using the empirical formulas \cite{Str82,Bla95} relating the high-energy
ISGDR peak in $^{208}$Pb and nuclear matter incompressibility $K_\infty$ we
obtain $K_\infty\approx 210$ MeV which is smaller than that deduced from the
ISGMR data by about 20 MeV.

The inelastic \aPb scattering data at $E_{\rm lab}=240$ MeV measured for the
energy bin centered at $E_x = 22.5$ MeV \cite{You04} and DFM + DWBA results
given by the CM transition densities scaled to the isoscalar $EL$ strengths
taken from Refs.~\cite{You04,Uch03,Uch03t} are shown in Fig.~\ref{f7}. After
substraction of the continuum contribution, the MDA \cite{You04} implied that
the measured \aap cross section (see Fig.~\ref{f7}) contains mainly the
isoscalar $E1$ and $E3$ components which exhaust, respectively, about 8\% and
6\% of the corresponding EWSR (see Table~\ref{t2}). The DFM + DWBA calculation
based exactly on these $E1$ and $E3$ transition strengths accounts reasonably
for the data (lower panel of Fig.~\ref{f7}), with some underestimation of the
data points at large angles (due perhaps to the contribution from the isovector
$E1$ mode, see Fig.~5 of Ref.~\cite{You04}). In contrast to the MDA results for
the 240 MeV data, in addition to similar isoscalar $E1$ and $E3$ strengths
around the peak $E_x = 22.5$ MeV, the MDA of the 386 MeV data \cite{Uch03,Uch04}
has found significant contributions from the isoscalar $E0$ and $E2$ strengths
which exhaust, respectively, about 2\% and 10\% of the corresponding EWSR
(Table~\ref{t2}). The DFM + DWBA calculation based on the isoscalar $EL$
strengths given by the MDA of the 386 MeV data also describes reasonably the 240
MeV data for the peak $E_x = 22.5$ MeV (upper panel of Fig.~\ref{f7}). The
high-energy tails of the $E0$ and $E2$ strength distributions given by the MDA
of the 386 MeV data give rise to an enhancement of the DWBA cross section at the
forward angles as shown in upper panel of Fig.~\ref{f7}. In Fig.~\ref{f8} the
386 MeV data measured for the 1 MeV bin centered at $E_x = 22.5$ MeV are
compared with the DFM + DWBA results given by the CM transition densities
(\ref{fd8})-(\ref{fd9}) based on the same isoscalar $EL$ transition strengths as
those used in Fig.~\ref{f7}. Similar to the DFM + DWBA results shown in
Figs.~\ref{f3} and \ref{f4}, the DFM + DWBA results for the peak $E_x = 22.5$
MeV strongly underestimate the 386 MeV data over the whole angular range.
However, the gap between the calculated and measured cross sections becomes
significantly larger in this case which shows the important contributions by the
excitation modes of higher multipoles ($L>3$) at energies above 20 MeV.

Given a sizable difference between the DFM + DWBA results and the MDA results of
Ref.~\cite{You04} found for the energy bin centered at the ISGMR peak shown in
Fig.~\ref{f4}, it is necessary to check this effect also for the energy bins
centered at the ISGQR and ISGDR peaks. In Fig.~\ref{f8k}, the DFM + DWBA results
for the \aap cross sections at $E_{\rm lab}=386$ MeV (in the 1-MeV bins centered
at $E_x = 10.5$ and 22.5 MeV) are compared with the corresponding MDA results by
Uchida {\it et al.} \cite{Uch03,Uch04} that are based on the same $EL$
strengths. Although in logarithmic scale these two sets of calculated DWBA cross
sections look similar in shape and strength, the MDA cross sections are larger
than those given by the DFM + DWBA calculation by about $40\sim 60\%$ over the
whole angular range. For the energy bin centered at the ISGQR peak, this
difference is up to $\sim 180\%$ at the most forward angles and is due mainly to
the difference in the $E2$ cross sections. Since the Coulomb contribution of the
inelastic scattering FF is quite strong at the forward angles for the $E2$
excitation mode, such an unusually large difference in the $E2$ cross sections
could be due to the different treatments of the Coulomb inelastic scattering FF
in the two approaches. Namely, the Coulomb FF is evaluated in our DFM approach
microscopically by double-folding the proton parts of the $^4$He g.s. density
and $^{208}$Pb transition density with the Coulomb interaction, with both the
direct and exchange terms calculated by a method similar to that used for the
nuclear FF, whereas the widely used ansatz for the Coulomb FF in the MDA of \aap
data is to assume a simple macroscopic model-independent formula \cite{Hor95}
containing the electric transition rate $B(EL)$ of the considered state. To
probe this effect, we have made the single-folding calculation \cite{KhoSat97}
of the nuclear FF for these cases, using exactly the same effective $\alpha N$
and CM transition densities as those used in the MDA by Uchida {\it et al.}
\cite{Uch03,Uch04}. The single-folded nuclear FF were then used with the same
microscopic Coulomb FF as that used in the DFM + DWBA calculation to calculate
the \aap cross sections in the energy bins centered at 10.5 and 22.5 MeV and the
results are plotted in Fig.~\ref{f8k} as dash-dotted curves. One can see that
the large difference at forward angles in the cross sections given by the
single- and double-folding methods is reduced significantly, and at large angles
the cross sections given by the single-folded FF are very close to the MDA cross
sections. The results shown in Fig.~\ref{f8k} stress again the need to use the
accurate DFM in the MDA of the \aap data to deduce the realistic $EL$ transition
strengths.

In conclusion, the DFM + DWBA analysis of the inelastic \aPb scattering data at
$E_{\rm lab}=240$ MeV measured in the energy bins centered at the peaks of the
ISGMR, ISGQR and ISGDR in $^{208}$Pb, using the CM nuclear transition densities
for the $EL$ excitation with $L\leqslant 3$, agree qualitatively with the
original MDA of these data \cite{You04}. Given a sizable difference in the \aap
cross sections obtained with the single- and double-folding methods for the
inelastic scattering FF, the uncertainties in the $EL$ strengths deduced from
the MDA of the considered data \cite{You04,Uch03,Uch04} might be significantly
larger. Similar DFM + DWBA analysis of the inelastic \aPb scattering data at
$E_{\rm lab}=386$ MeV strongly underestimates the data points measured in about
the same energy bins and, thus, indicates a strong contribution by the
high-multipole ($L>3$) excitation modes. However, the gap between the calculated
and measured cross sections is quite different in the excitation energy regions
around 10 MeV and above 20 MeV. This result shows that the maximum angular
momentum $L_{\rm max}$ taken into account in the MDA series (\ref{MDA}) seems to
be energy dependent if the background due to the high-multipole excitation modes
is not explicitly subtracted which is the case for the 386 MeV data
\cite{You04,Uch03,Uch03t}.

\section{Results obtained with the microscopic RPA transition densities}
\label{sec4}

Although there is no consensus whether microscopic models like RPA can provide
reliable inputs for the nuclear transition densities, it has been shown in the
past \cite{khoapp,Kho86,Ho96} that for the low-lying excited states of
dominating one-phonon structure, the RPA transition densities can be
successfully used in the folding model analysis. The DFM + DWBA description of
inelastic \aPb scattering data measured for the $3^-_1$ state of $^{208}$Pb
given by the RPA nuclear transition density (see lower panel of Fig.~\ref{f1})
is again a convincing example. Of course, there are cases in which RPA is known
to have drawbacks, like in the case of low-lying states with a strong
\emph{anharmonic} mixture of the two-phonon structure \cite{Kho86}. Isoscalar
giant resonances, as already discussed above, should be a good test ground for
the RPA wave functions because RPA has been claimed over the years to be a
proper theory to describe those resonances. However, these qualitative arguments
are often invoked but in fact we are not aware of the conclusive evidences
showing that microscopic RPA provides accurate transition densities for
inelastic scattering calculations (cf., in this respect, Refs. \cite{Kol00}).
Also, this is probably the first combination of a {\em fully} self-consistent
RPA approach and an advanced microscopic double-folding model, and consequently
in the present context the question of the accuracy of the RPA transition
densities can be addressed carefully in more detail.

It is impossible to reproduce the full experimental width of a resonance state
within the RPA approach. Although the width caused by fragmentation of the
resonance strength (the so-called Landau damping) can be accounted for within
the RPA, and the escape width can also be accounted for if continuum-RPA is
performed, the spreading width (which is by far the most relevant in heavy
nuclei) cannot come out from RPA. To have a direct quantitative comparison of
the RPA solution with the observed $EL$ strength distribution, we found it
necessary to perform some \emph{averaging} \cite{Fe92} of the total RPA strength
(\ref{rpa2}) over the excitation energy $E_x$ as follows
\begin{equation}
 \left< S^{\rm RPA}_L(E_x) \right>
 = \sum_\nu S^{\rm RPA}_L(E_\nu)f(E_x-E_\nu), \label{aver}
\end{equation}
where $\nu$ labels the RPA (isoscalar $EL$) states and a Lorentzian \cite{Fe92}
is used as the averaging function $f(E-E')$. In each case, the averaging width
$\Delta$ is chosen so that the averaged RPA strength
\begin{equation}
 \int_{E_x-0.5\ {\rm MeV}}^{E_x+0.5\ {\rm MeV}} \left< S^{\rm RPA}_L(E) \right>
 dE \label{aver1}
\end{equation}
within the 1-MeV energy bin centered at an excitation energy $E_x$ is smooth
enough to be compared with that deduced from the MDA analysis of \aap data. In
the case of ISGMR, a strongly collective resonance RPA state was found at
$E_x\approx 14.2$ MeV which is quite close to the experimental ISGMR peak. This
RPA state is accompanied by several non-collective RPA states on either sides of
the peak (see upper panel of Fig.~\ref{f9}) and the whole set of monopole RPA
states exhausts about 99.5\% of the $E0$ EWSR. The averaging width
$\Delta\approx 3$ MeV, which is close to the observed ISGMR width \cite{You04},
was chosen to smooth the strength of the RPA resonance state over the excitation
energy. The distribution of \emph{averaged} RPA strength agrees reasonably with
those deduced from the MDA analyses of the \aap data at 240 MeV \cite{You04} and
386 MeV \cite{Uch03,Uch03t} as shown in the lower panel of Fig.~\ref{f9}.

For the ISGQR, most of the microscopic RPA calculations predict a strong $E2$
resonance at somewhat higher excitation energy ($E_x\approx 12.5$ MeV in our
case) compared to the experimental ISGQR peak around 10.3 MeV, in keeping with
the low effective mass associated with effective Skyrme or Gogny interactions.
The lowest $2^+_1$ state is predicted by the RPA at an excitation energy of
$E_x\approx 5.1$ MeV which is also higher than the experimental value of about
4.09 MeV. To have comparable $E2$ strengths at the ISGQR peak observed in \aap
experiment, we have shifted all isoscalar quadrupole RPA states (which exhaust
99.3\% of the $E2$ EWSR) down by 2 MeV in the excitation energy and the resulted
RPA spectrum is shown in upper panel of Fig.~\ref{f10}. By using the same
averaging width $\Delta\approx 3$ MeV which is also close to the observed ISGQR
width \cite{You04}, the distribution of averaged RPA strength agrees well with
those deduced from the MDA analyses of the \aap data at both energies (see lower
panel of Fig.~\ref{f10}). While the averaged RPA strength and that given by the
MDA of 240 MeV data \cite{You04} are concentrated mainly near the ISGQR peak and
slightly spread over $E_x\leq 21$ MeV, the $E2$ strength given by the MDA of 386
MeV data \cite{Uch03,Uch03t} is broadly spread up to much higher excitation
energies.

Concerning the ISGDR excitation in $^{208}$Pb, most of the microscopic structure
calculations \cite{Col00,Vre00,Pie01} predict the main (high-energy) peak of the
ISGDR at $E_x\approx 24.5\sim 25.5$ MeV which is somewhat higher than that
($E_x\approx 22.5$ MeV) observed in the \aap experiments
\cite{You04,Uch03,Uch03t}. Like the earlier RPA results obtained with the SLy4
interaction \cite{Col00}, the present RPA calculation using the SLy5 interaction
predicts the high-energy peak of the ISGDR at $E_x\approx 25$ MeV. In a similar
manner, we have shifted all the isoscalar dipole RPA states (which exhaust
86.5\% of the $E1$ EWSR) down by 3 MeV in excitation energy and the shifted
spectrum of dipole RPA states is shown in the upper panel of Fig.~\ref{f11}. To
have a better resolution of the averaged dipole strength of numerous RPA states
found in the resonance region, we have used a finer averaging width of
$\Delta\approx 1$ MeV in this case and the distribution of averaged RPA strength
agrees reasonably with the observed dipole strength distribution at the main
ISGDR peak (see lower panel of Fig.~\ref{f11}). However, as it can be seen in
the lower panel of Fig.~\ref{f11}, the low-energy ISGDR strength observed at
$E_x\approx 13$ MeV in both \aap experiments \cite{You04,Uch03,Uch03t} is not
reproduced by the present RPA calculation. Whether this is an experimental or
theoretical problem, is an open question. Certainly in most of the microscopic
RPA calculations the low-energy ISGDR strength is less collective than the
high-energy part (see, e.g., the discussion in Ref. \cite{Col00}) and, as such,
less amenable to a RPA description. Suggestions that this strength corresponds
to toroidal motion have been put forward \cite{toroidal} and if this were the
case, the capability of microscopic functionals to describe such exotic mode,
and the relationship with \aap cross sections is even less clear.

Given the energy distributions of the $E0,\ E1$ and $E2$ strengths reasonably
described by the \emph{averaged} RPA strengths as shown above, it is natural to
expect that the DFM + DWBA calculation using a proper input of the RPA
transition densities should also deliver a good description of the measured \aap
cross sections. However, to compare the DFM + DWBA results with the \aap cross
section measured for a given energy bin, one needs to combine properly the
transition densities of all RPA states in this energy bin into a total RPA
transition density which can be used as input of the DFM calculation. In
general, if the number of RPA solutions for a given $EL$ transition in the
energy bin $dE$ centered at $E_x$ is $N$, then the total RPA transition density
associated with the bin $dE$ should be defined as
\begin{equation}
\delta\rho^{\rm RPA}_{L}(r)=\sum_{\nu=1}^N\ M_{L\nu}\delta\rho_{L\nu}(r);
 \ {\rm with}\ S^{\rm RPA}_L=\sum_{\nu=1}^NS^{\rm RPA}_{L\nu}
 \equiv \sum_{\nu=1}^N|M_{L\nu}|^2.
 \label{denav1}
\end{equation}
Here $M_{L\nu}$ is the RPA transition moment (\ref{rpa2}) of the RPA state
$|L\nu>$ and $\delta\rho_{L\nu}(r)$ is the corresponding RPA transition density
(\ref{rpa1}). It is natural to choose the transition moment $M_{L\nu}$ as the
averaging weight for the RPA transition density $\delta\rho_{L\nu}(r)$, so that
the RPA transition density (\ref{denav1}) preserves the total transition
strength $S^{\rm RPA}_L$ in the considered energy bin as predicted by the RPA
calculation. The only question now is whether the total RPA transition density
should be scaled to reproduce the averaged RPA strength (\ref{aver1}) in this
energy bin or should it be kept unchanged as given by the weighted sum
(\ref{denav1}) of all RPA transition densities. We have found, however, that the
first procedure is reasonable only for the transition densities
$\delta\rho_{L\nu}(r)$ of the strongest RPA states with $L=0$ and 2 in the
energy bins around the ISGMR and ISGQR peaks, respectively. For the RPA states
of other multipolarities ($L=1,3$) in the same energy bins the RPA spectrum
consists only of a few non-collective states and such a scaling procedure can
lead to very unrealistic shapes of $\delta\rho_{L\nu}(r)$ with $L=1$ and 3 which
strongly distort the calculated \aap cross section for the ISGMR and ISGQR
peaks. We have used, therefore, the total RPA transition density as given by the
weighted sum (\ref{denav1}) of all RPA transition densities in the considered
energy bin for the DFM + DWBA calculation of the \aap cross section.

To compare our DFM + DWBA results with the inelastic \aPb scattering data at
$E_{\rm lab}=240$ MeV \cite{You04}, we have generated the total RPA transition
densities (\ref{denav1}) for the three 640-keV energy bins centered at $E_x =
10.3$, 14.1 and 22.5 MeV, respectively. The total RPA transition density and the
two strongest individual RPA transition densities (\ref{rpa1}) in each energy
bin are compared in Fig.~\ref{f12} with the total collective model transition
density (\ref{fd6})-(\ref{fd11}) based on $EL$ strengths given by the MDA of the
240 MeV data \cite{You04}. While the radial shape of the total RPA transition
densities in the energy bins centered at the ISGQR and ISGMR peaks agrees more
or less with that of the CM densities, the $EL$ strengths given by the RPA
transition densities are much stronger than those given by the CM densities.
Such an effect is not unexpected because the predicted $E0$ and $E2$ strengths
near the ISGQR and ISQMR peaks are concentrated in just a few discrete RPA
states (see upper panels of Figs.~\ref{f9} and \ref{f10}). In contrary, the RPA
strength for the isoscalar dipole excitation are distributed over many weakly
excited $E1$ states, and the total RPA dipole transition density given by the
weighted sum (\ref{denav1}) of 15 RPA states in the 640-keV bin centered at
$E_x=22.5$ MeV has a slightly weaker $E1$ strength compared to that of the CM
transition density. In other words, the RPA calculations of the $E0$ and $E2$
strength do not predict fragmentation, at variance with the $E1$ case where the
fragmentation caused by Landau damping is quite large. The $EL$ transition
strengths given by the total RPA transition densities (\ref{denav1}) in the
three 640-keV energy bins centered at the ISGMR, ISGQR and ISGDR peaks are given
in Table~\ref{t3}.

The DFM + DWBA description of the \aap data at 240 MeV \cite{You04} given by the
RPA transition densities are shown in Fig.~\ref{f13}. Although these DFM + DWBA
results agree well with the data, in about the same way as the results given by
the CM transition densities, the $EL$ strengths associated with the RPA
transition densities in each energy bin are quite different from those given by
the CM transition densities (see Table~\ref{t2}). In energy bins centered at the
ISGMR and ISGDR peaks the $EL$ transition strengths with $L=0$ and 2,
respectively, are much more dominant compared to those deduced from the MDA of
the \aap data. For example, only a single RPA transition density (\ref{rpa1}) of
the strongest (discrete) $2^+$ state at the ISGQR peak accounts perfectly
for the data measured for the energy bin centered at $E_x=10.3$ MeV. However,
this result should not give the wrong impression that the $EL$ strengths from
other multipoles are not as significant as given by the MDA of 240 MeV data
\cite{You04}. In reality, the isoscalar giant resonances under study are not
discrete states as predicted by the RPA but widely fragmented over the
excitation energy, having widths $\Gamma\approx 3\sim 4$ MeV. The physical
origin of the observed $EL$ strength fragmentation of the isoscalar giant
resonances, in particular their spreading widths, can be described by invoking
the anharmonic effects beyond RPA caused by, e.g., the coupling with 2p-2h type
configurations (see Refs. \cite{Ber83,Wam90} for extensive reviews).
The inclusion of the anharmonic effects is expected not only to
redistribute the strength of the resonance RPA state over the energy like the
averaging procedure (\ref{aver}), but also to pull down the predicted excitation
energy to a lower value. We recall again here that the RPA spectrum of $2^+$
states has been shifted down by 2 MeV in energy to have the strongest $2^+$
state near the observed ISGQR peak (see upper panel of Fig.~\ref{f10});
this shift is precisely associated with 2p-2h coupling or in other words, as
stated above, to the renormalization of the effective mass $m^*$.
Nevertheless, such a perfect agreement of the DWBA cross section predicted by
the total RPA transition density with the data measured for the ISGQR peak (see
upper panel of Fig.~\ref{f13}) without any fine-tuning of the FF strength is
very encouraging. This result confirms the realistic shape of the RPA transition
density shown in Fig.~\ref{f12} and stresses once more the strong predicting power
of the RPA approach.

For the ISGMR, the DFM + DWBA calculation using a single transition density
(\ref{rpa1}) for the collective RPA state near the ISGMR peak predicts values of
the \aap cross section quite close to the data in magnitude but with a deep
oscillation pattern that can be smoothed out only by adding the contributions
from other multipoles. Such a behavior of the \aap cross section calculated for
the ISGMR excitation in $^{208}$Pb has been seen in the earlier folding model
studies \cite{Hor95,Ber80}. Although the non-zero multipole $EL$ transition
strengths predicted by the RPA for the energy bin centered at $E_x=14.1$ MeV are
somewhat weaker than those given by the MDA of 240 MeV data (see Tables~\ref{t2}
and \ref{t3}), their contributions are still essential in smoothing out the deep
oscillation of the DWBA cross section predicted by the $E0$ transition. Given
the quite accurate RPA description of both the $E0$ strength distribution (upper
panel of Fig.~\ref{f9}) and of the \aap cross section measured for the ISGMR
peak (middle panel of Fig.~\ref{f13}) without any readjustment of the model
parameters and shift of the excitation energy, we conclude that the RPA is
indeed a reliable theoretical approach to study the GMR excitations in nuclei.

For the ISGDR, the DFM + DWBA calculation using the total RPA transition
densities (\ref{rpa1}) for $L\leq 3$ in the energy bin centered at $E_x= 22.5$
MeV reproduces very well the \aap angular distribution measured at 240 MeV (see
lower panel of Fig.~\ref{f13}). The main difference between these results and
the DFM + DWBA results given by the CM transition densities shown in
Fig.~\ref{f7} is that the RPA predicts quite a strong $E3$ strength in this
energy bin ($\sim 10.8\%$) compared to that deduced from the MDA of 240 MeV data
($\sim 5.9\%$), while the $E1$ strength predicted by the RPA is about 60\%
weaker than the MDA value. As a result, the $E3$ transition turned out to give a
dominant contribution to the DWBA cross section for the energy bin centered at
$E_x= 22.5$ MeV.

In conclusion, our study of the ISGMR, ISGQR and ISGDR strength distributions
and inelastic scattering \aap angular distributions measured at the
corresponding resonance peaks using the RPA transition densities has shown that
the RPA approach can be successfully used to describe not only the $EL$ strength
distribution and integral properties of the giant resonances, but also the
(double) differential $d^2\sigma/d\Omega dE$ cross sections. Although there are
some differences in the $EL$ strengths of non-collective states, the ISGMR,
ISGQR and ISGDR strength distributions predicted by the \emph{averaged} RPA
results agree reasonably good with those deduced from the MDA of the \aap data
(see Table~\ref{t2} and Figs.~\ref{f9}, \ref{f10} and \ref{f12}). The $EL$
transition strengths as well as the location of the ISGQR and ISGDR peaks
originally predicted by the RPA are significantly higher than those given by the
MDA of the considered \aap data. This effect can be qualitatively explained by
the lack of the (beyond RPA) anharmonic contributions from, e.g., the 2p-2h
coupling.

Therefore, it would be of further interest to perform the same DFM + DWBA
analysis including effects beyond the RPA since a self-consistent microscopic
calculation beyond RPA coupled to an accurate reaction framework does not exist
yet, to our knowledge. The effects, which have been introduced {\em ad hoc} in
the present work via the shift of the RPA mean peak(s) downwards and spreading
(\ref{aver})-(\ref{aver1}) of the discrete RPA strength, are exactly those
included microscopically into a second RPA, or RPA plus phonon coupling (RPA-PC)
calculations (see also Ref.~\cite{Kam09}). In this sense, the present results
are encouraging. In particular, the effects of 2p-2h coupling on the ISGMR are
expected to be small due to well-known cancellation effects and this goes along
with the fact that our present simple RPA gives a very good DFM + DWBA
description of the monopole excitation.

\section{Summary}

The generalized double-folding approach of Refs.~\cite{Kho00,Kho97} has been
further developed for the microscopic study of isoscalar giant resonances
induced by inelastic \aA scattering, using the nuclear transition densities
given by both the collective model and microscopic self-consistent RPA
calculation. Although the single-folding approach \cite{KhoSat97} has been often
used to compute the \aA potential in the MDA of the \aap data, we have shown in
the present work a significant difference in the \aap cross sections given by
the single- and double-folding methods. The DFM has also been suggested earlier
\cite{KhoSat97,Hor95} as a more accurate approach to obtain realistic results
for the inelastic \aA and HI scattering, in particular, the EWSR fractions
exhausted by different $EL$ excitation modes. Therefore, the present combination
of the DFM and DWBA approaches should be a reliable alternative method to be
used in the MDA of future \aap measurements.

A reasonable DFM + DWBA description of the 240 MeV \aap data in energy bins
centered at the ISGMR, ISGQR and ISGDR peaks in $^{208}$Pb has been obtained
with the CM nuclear transition densities built upon the $EL$ strengths given by
the MDA of these data. Our similar study of the inelastic \aPb scattering at
$E_{\rm lab}=386$ MeV strongly underestimates the data points measured in the
same energy bins and, thus, shows quite a strong contribution by the excitation
modes of higher multipole ($L>3$). The contribution by higher multipole
excitation modes was found significantly different in the two energy regions
around 10 MeV and above 20 MeV, which suggests that the maximum angular momentum
$L_{\rm max}$ taken into account in the MDA series (\ref{MDA}) should be energy
dependent.

The present DFM + DWBA method has also provided an accurate direct link between
the \emph{discrete} RPA approach (which was used in the past to describe mainly
the integral properties of the giant resonances) and the experimental double
differential \aap cross sections measured for the resonance peaks. Given
high-precision \aap data for the isoscalar giant resonances under study, our
method can be used in the future to probe the accuracy of the microscopic
prediction by different structure models for the energy distribution of the
$S_L(E)$ strength as well as the inelastic scattering $d^2\sigma/d\Omega dE$
cross section. In such a connection, we would like to emphasize again that the
latter is very sensitive to the interference of different $EL$ contributions.

\section*{Acknowledgments}
We are indebted to Prof. D.H. Youngblood, Prof. T. Kawabata, Dr. X. Chen and Dr.
M. Uchida for their helpful communications on the \aap data under study. The
present research has been supported by National Foundation for Scientific \&
Technological Development (NAFOSTED) under Project No 103.04.07.09. D.C.C.
gratefully acknowledges the financial support from the Asia Link Programme
CN/Asia-Link 008 (94791) during his two research stays at the University of
Milan where this work has been initiated. Last but not least, one of us (D.T.K.)
is grateful to Ray Satchler for his numerous discussions during the years of
collaboration, which have laid the foundation for this research study.

\newpage
 \centerline{\bf Figure captions}
\bigskip

Fig.1: Upper panel: OM description of the elastic \aPb scattering data at 240
MeV \cite{You04} and 386 MeV \cite{Uch03,Uch04} given by the unrenormalized
(dash curves) and renormalized complex folded OP (solid curves). Lower panel:
DWBA descriptions of the inelastic \aPb scattering data \cite{You04,Uch03,Uch04}
for $3^-_1$ state of $^{208}$Pb ($E_x = 2.61$ MeV) given by the collective model
(dash curves) and RPA (solid curves) nuclear transition densities.

Fig.2: Inelastic \aPb scattering data at $E_{\rm lab}=240$ MeV measured for the
energy bin centered at $E_x = 14.1$ MeV \cite{You04} in comparison with the DFM
+ DWBA results given by the CM transition densities based on the isoscalar $EL$
strengths taken from Refs.~\cite{You04} (lower panel) and \cite{Uch03} (upper
panel). See details in the text and Table~\ref{t2}.

Fig.3: Inelastic \aPb scattering data at $E_{\rm lab}=386$ MeV measured for the
energy bin centered at $E_x = 13.5$ MeV \cite{Uch03t} in comparison with the DFM
+ DWBA results given by the CM transition densities based on the isoscalar $EL$
strengths taken from Refs.~\cite{You04} (lower panel) and \cite{Uch03,Uch04}
(upper panel). See details in the text and Table~\ref{t2}.

Fig.4: Inelastic \aPb scattering data at $E_{\rm lab}=386$ MeV measured for the
energy bin centered at $E_x = 14.5$ MeV \cite{Uch03t} in comparison with the
DWBA results given by the isoscalar $EL$ strengths taken from
Refs.~\cite{Uch03,Uch04}. Upper panel: the present DFM + DWBA calculation; lower
panel: the MDA results by Uchida {\it et al.} \cite{Uch03,Uch04}.

\begin{figure}[bht]
 \begin{center}\vspace{-2cm}\hspace{-1cm}
\includegraphics[angle=0,scale=0.70]{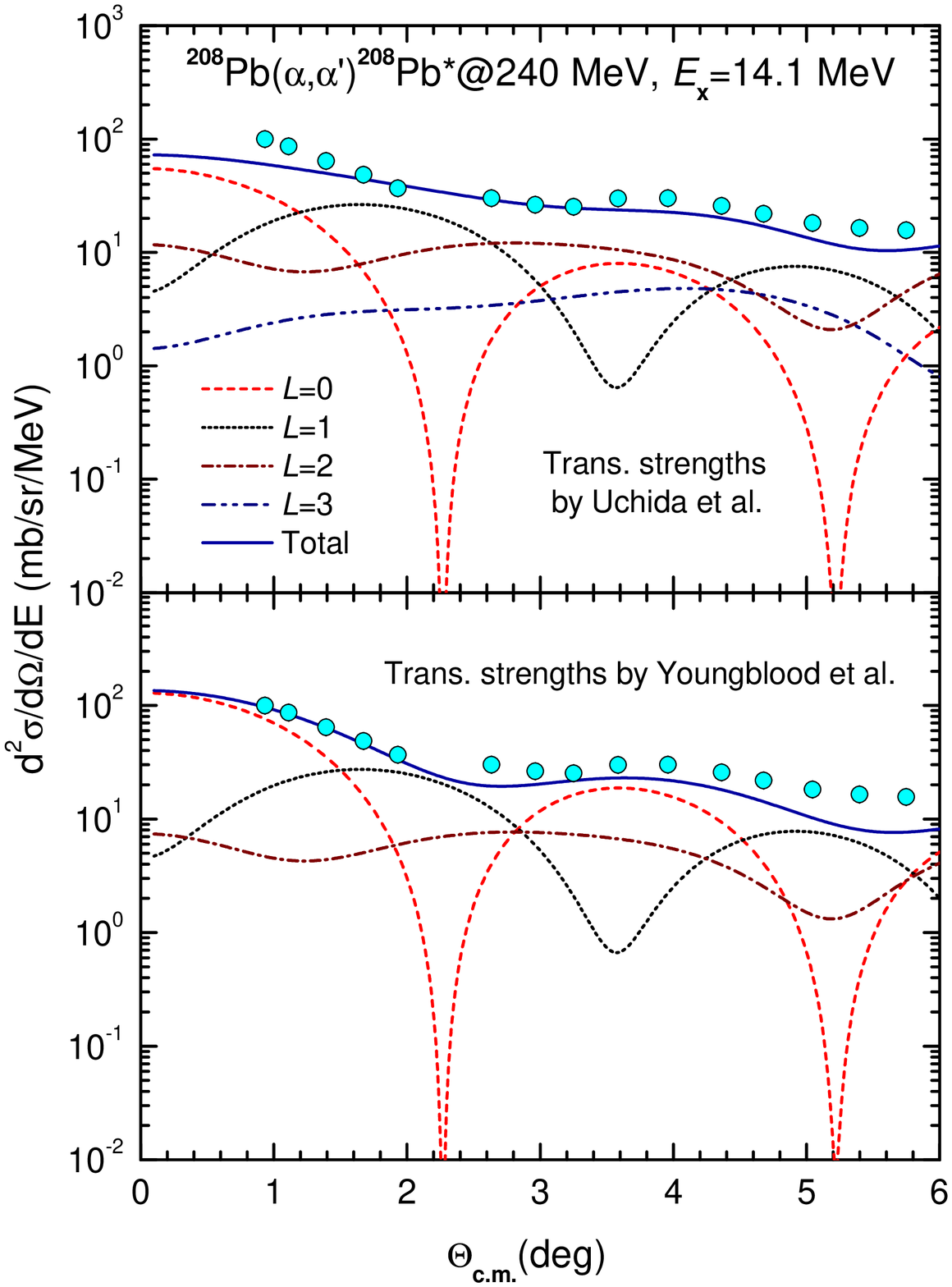}\vspace{-1cm}
\caption{Inelastic \aPb scattering data at $E_{\rm lab}=240$ MeV measured for
the energy bin centered at $E_x = 14.1$ MeV \cite{You04} in comparison with the
DFM + DWBA results given by the CM transition densities based on the isoscalar
$EL$ strengths taken from Refs.~\cite{You04} (lower panel) and \cite{Uch03}
(upper panel). See details in the text and Table~\ref{t2}.}\label{f2}
\end{center}
\end{figure}

Fig.5: The same as Fig.~\ref{f2} but for the energy bin centered at
 $E_x=10.3$ MeV.

\begin{figure}[bht]
 \begin{center}\vspace{-2cm}\hspace{-1cm}
\includegraphics[angle=0,scale=0.70]{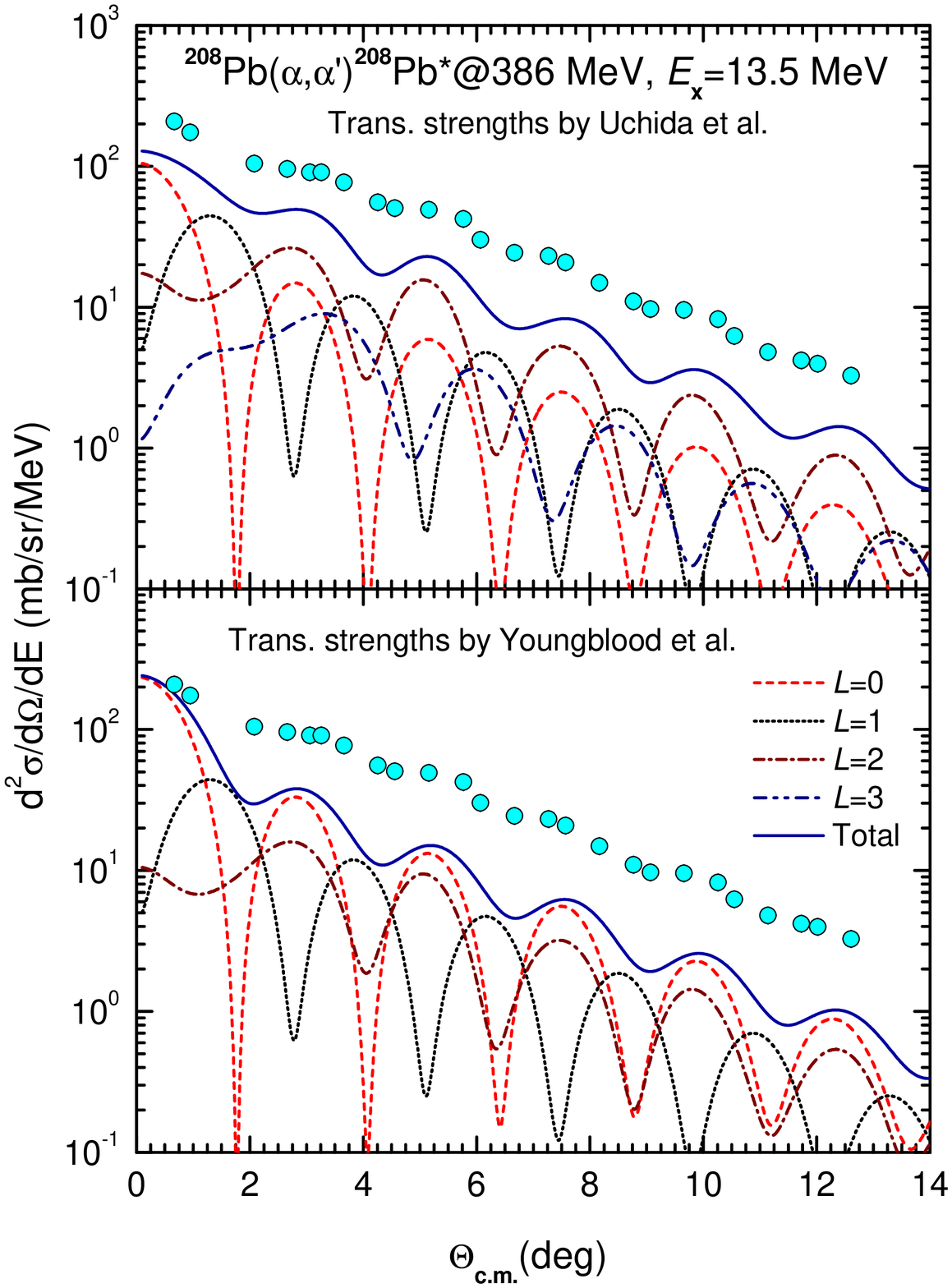}\vspace{-1cm}
\caption{Inelastic \aPb scattering data at $E_{\rm lab}=386$ MeV measured for
the energy bin centered at $E_x = 13.5$ MeV \cite{Uch03t} in comparison with the
DFM + DWBA results given by the CM transition densities based on the isoscalar
$EL$ strengths taken from Refs.~\cite{You04} (lower panel) and
\cite{Uch03,Uch04} (upper panel). See details in the text and
Table~\ref{t2}.}\label{f3}
\end{center}
\end{figure}

\begin{figure}[bht]
 \begin{center}\vspace{-2cm}\hspace{-1cm}
\includegraphics[angle=0,scale=0.70]{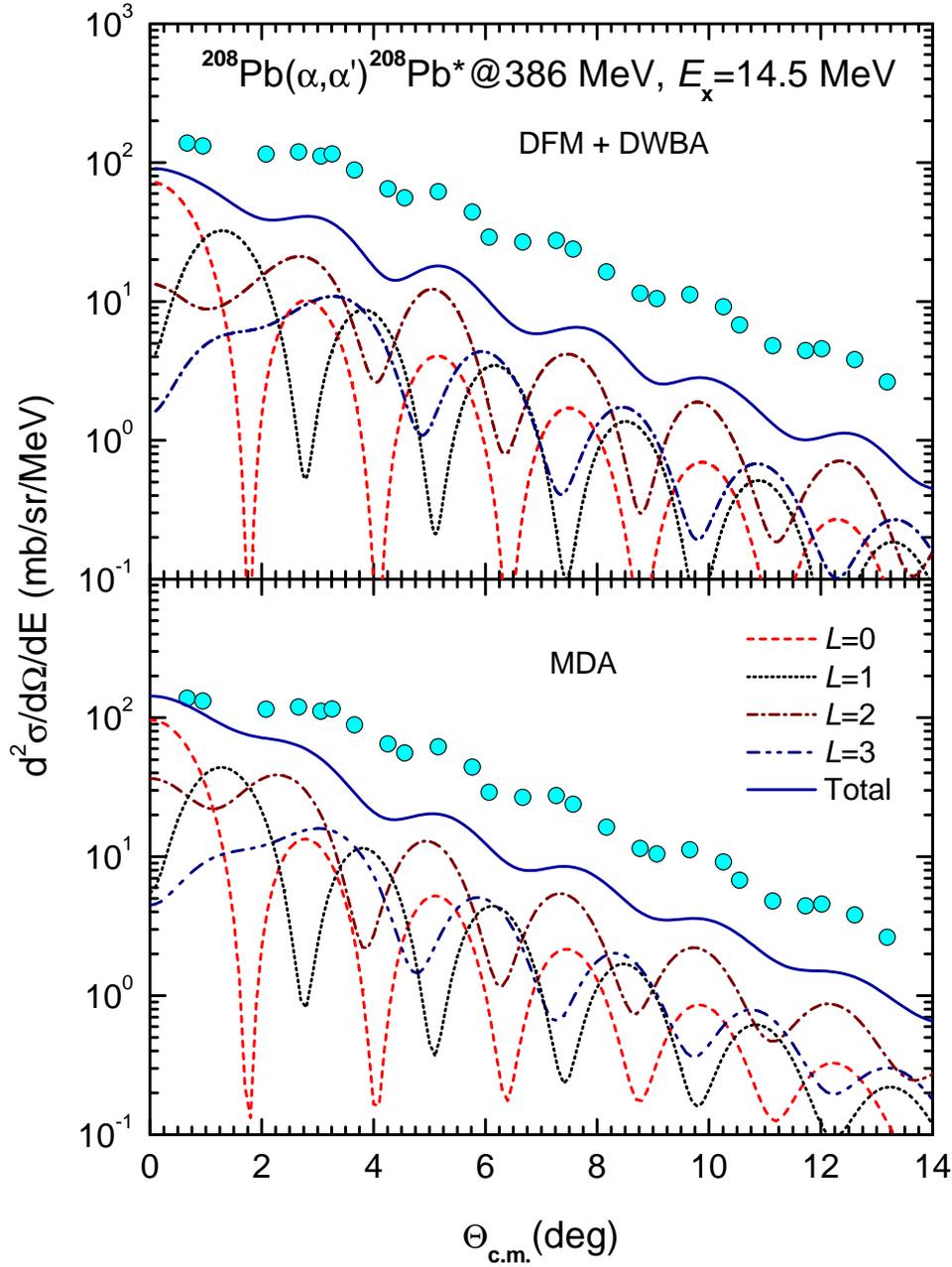}\vspace{-1cm}
\caption{Inelastic \aPb scattering data at $E_{\rm lab}=386$ MeV measured for
the energy bin centered at $E_x = 14.5$ MeV \cite{Uch03t} in comparison with the
DWBA results given by the isoscalar $EL$ strengths taken from
Refs.~\cite{Uch03,Uch04}. Upper panel: the present DFM + DWBA calculation; lower
panel: the MDA results by Uchida {\it et al.} \cite{Uch03,Uch04}.}\label{f4}
\end{center}
\end{figure}

\begin{figure}[bht]
\begin{center}\vspace{-2cm}\hspace{-1cm}
\includegraphics[angle=0,scale=0.70]{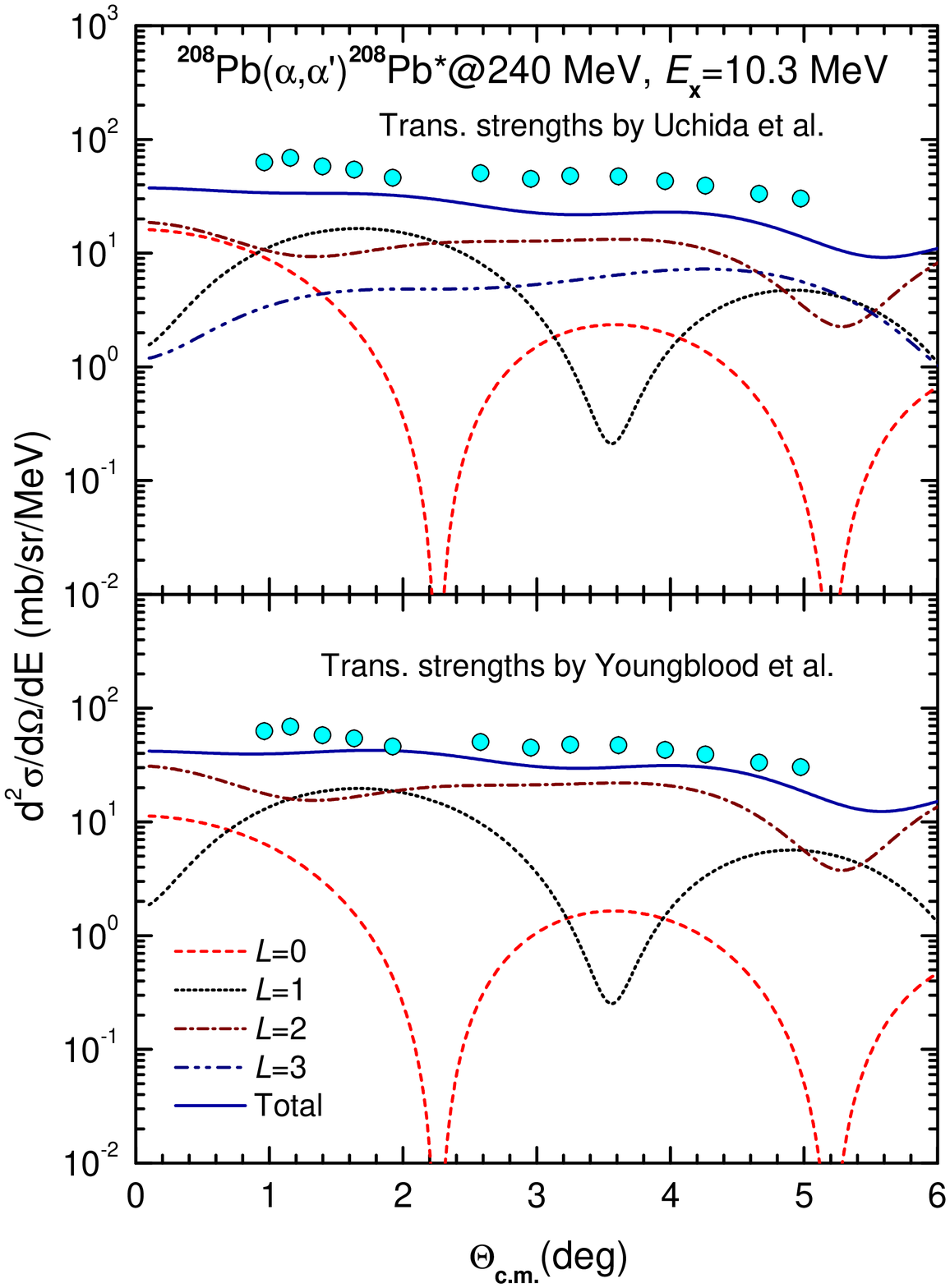}\vspace{-1cm}
\caption{The same as Fig.~\ref{f2} but for the energy bin centered at
 $E_x=10.3$ MeV.}\label{f5}
\end{center}
\end{figure}

\begin{figure}[bht]
\begin{center}\vspace{-2cm}\hspace{-1cm}
\includegraphics[angle=0,scale=0.70]{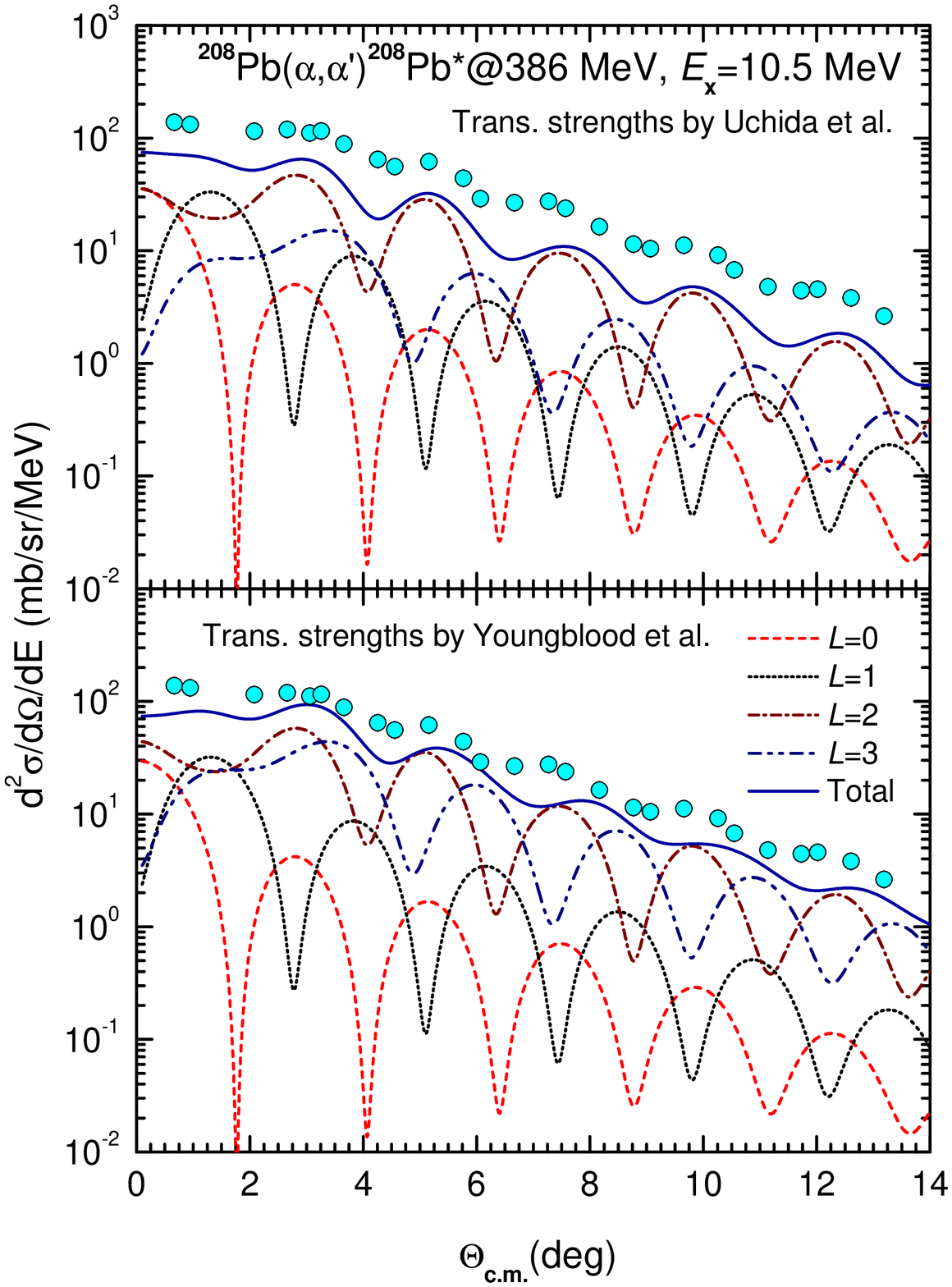}\vspace{-1cm}
\caption{The same as Fig.~\ref{f3} but for the energy bin centered at
 $E_x=10.5$ MeV.}\label{f6}
\end{center}
\end{figure}

\begin{figure}[bht]
\begin{center}\vspace{-2cm}\hspace{-1cm}
\includegraphics[angle=0,scale=0.70]{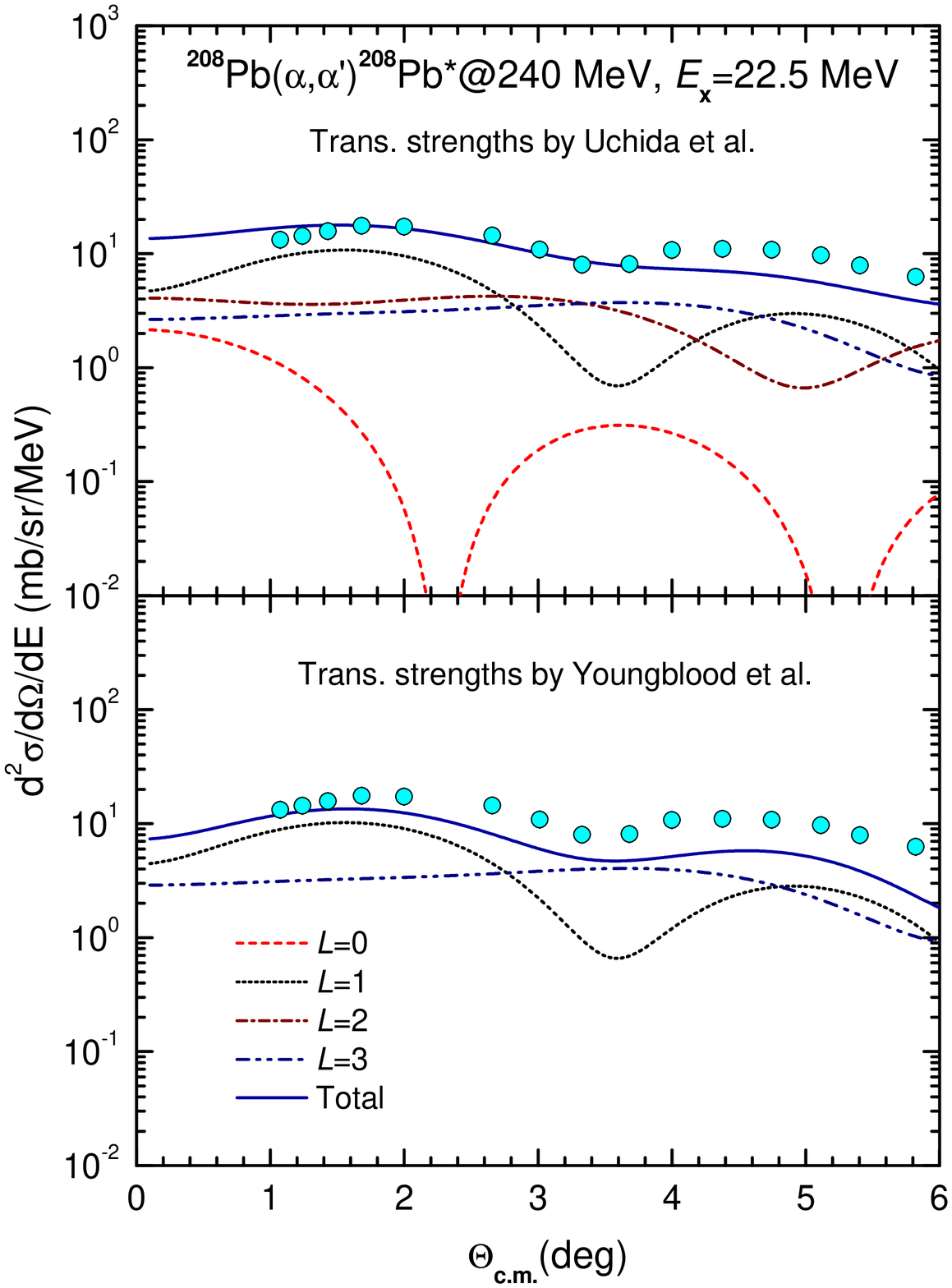}\vspace{-1cm}
\caption{The same as Fig.~\ref{f2} but for the energy bin centered at
 $E_x=22.5$ MeV.}\label{f7}
\end{center}
\end{figure}

\begin{figure}[bht]
\begin{center}\vspace{-2cm}\hspace{-1cm}
\includegraphics[angle=0,scale=0.70]{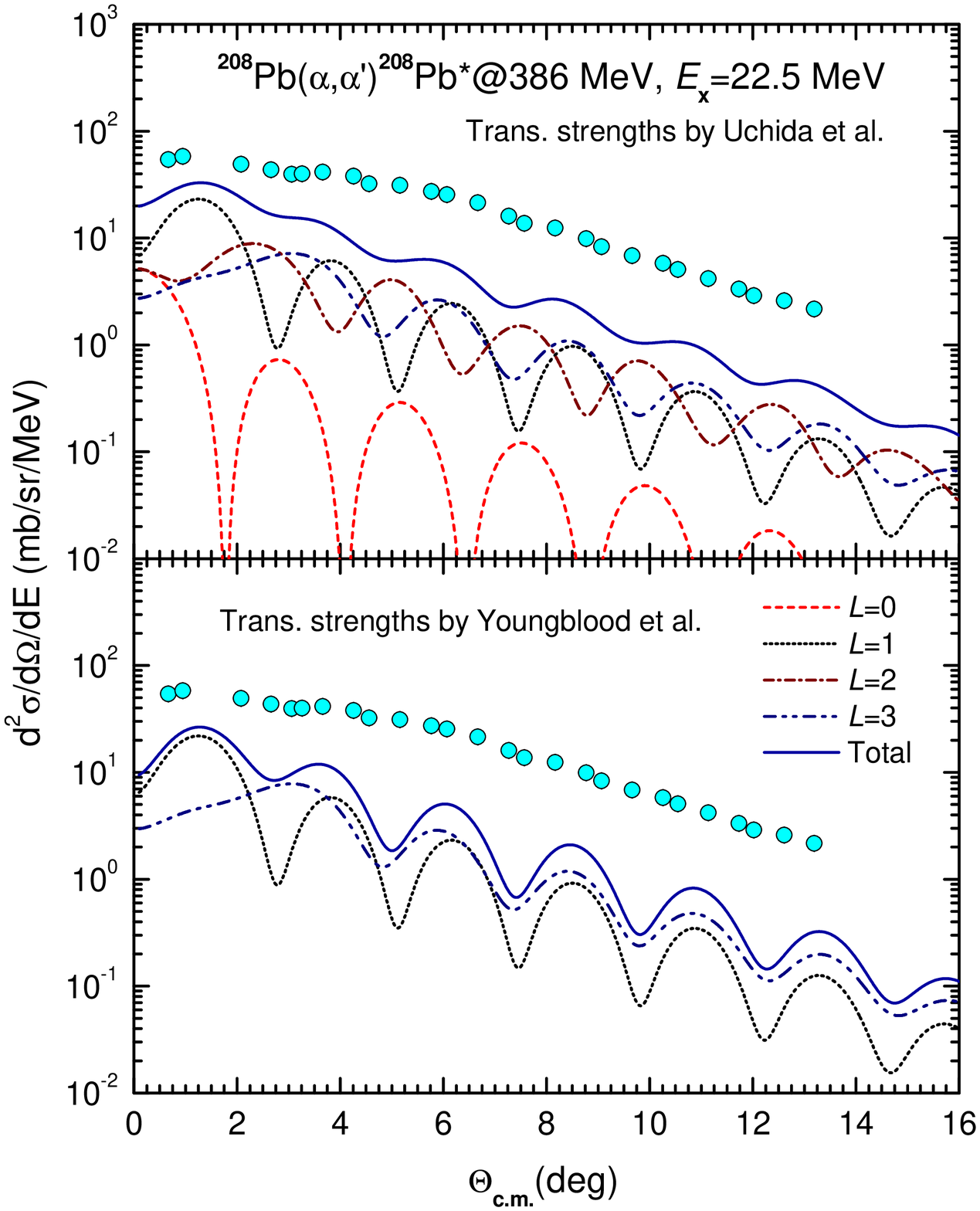}\vspace{-1cm}
\caption{The same as Fig.~\ref{f3} but for the energy bin centered at
 $E_x=22.5$ MeV.}\label{f8}
\end{center}
\end{figure}

\begin{figure}[bht]
 \begin{center}\vspace{-2cm}\hspace{-1cm}
\includegraphics[angle=0,scale=0.70]{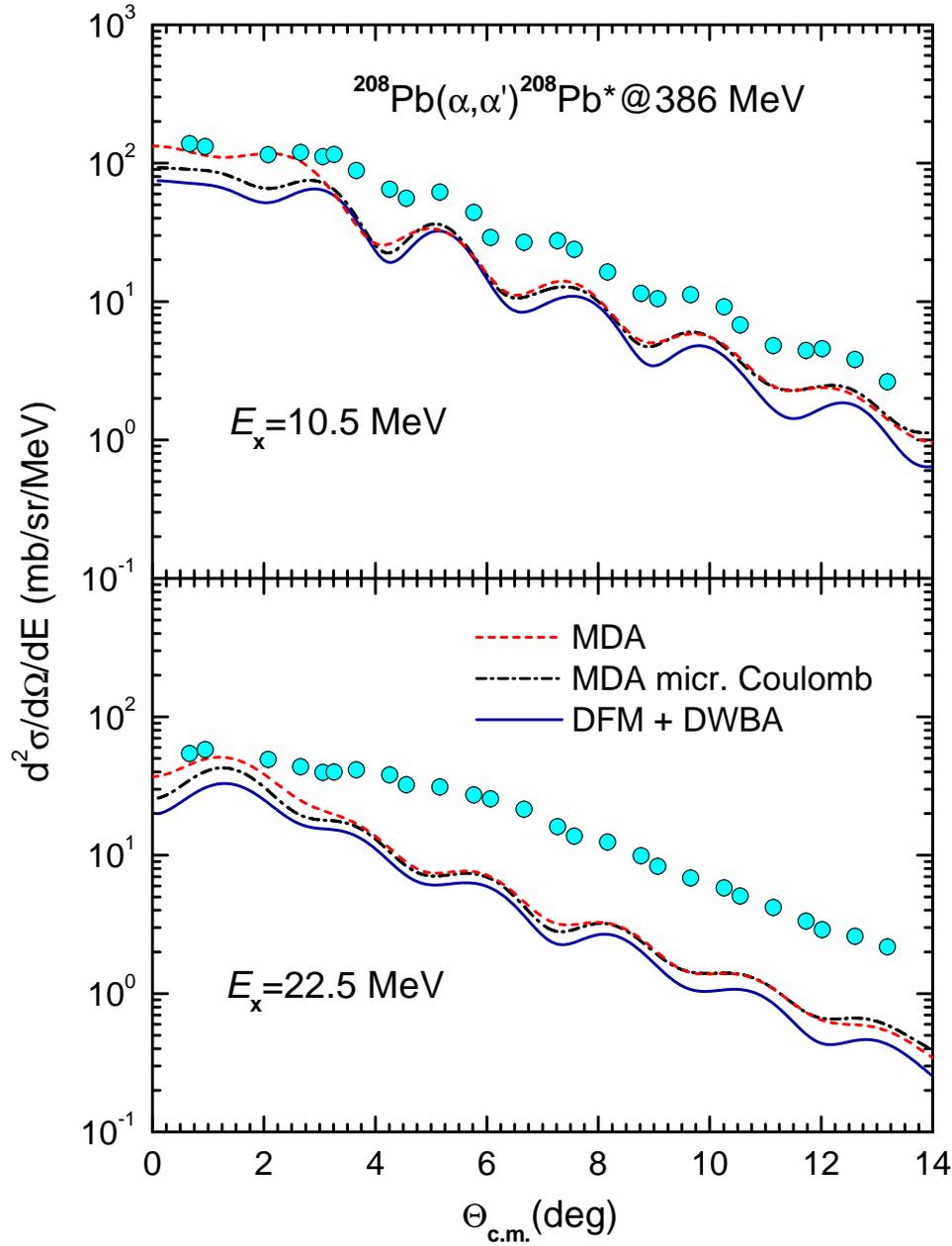}\vspace{-1cm}
\caption{Inelastic \aPb scattering data at $E_{\rm lab}=386$ MeV measured for
the energy bins centered at $E_x = 10.5$ and 22.5 MeV \cite{Uch03t} in
comparison with the DWBA results given by the isoscalar $EL$ strengths taken
from Refs.~\cite{Uch03,Uch04}. Solid curves: the present DFM + DWBA calculation;
dashed curves: the MDA results by Uchida {\it et al.} \cite{Uch03,Uch04};
dashed-dotted curves: the same as dashed curves but using microscopic Coulomb FF
from the present DFM calculation.}\label{f8k}
\end{center}
\end{figure}

Fig.6: The same as Fig.~\ref{f3} but for the energy bin centered at
 $E_x=10.5$ MeV.

Fig.7: The same as Fig.~\ref{f2} but for the energy bin centered at
 $E_x=22.5$ MeV.

Fig.8: The same as Fig.~\ref{f3} but for the energy bin centered at
 $E_x=22.5$ MeV.

Fig.9: Inelastic \aPb scattering data at $E_{\rm lab}=386$ MeV measured for the
energy bins centered at $E_x = 10.5$ and 22.5 MeV \cite{Uch03t} in comparison
with the DWBA results given by the isoscalar $EL$ strengths taken from
Refs.~\cite{Uch03,Uch04}. Solid curves: the present DFM + DWBA calculation;
dashed curves: the MDA results by Uchida {\it et al.} \cite{Uch03,Uch04};
dashed-dotted curves: the same as dashed curves but using microscopic Coulomb FF
from the present DFM calculation.

Fig.10: Isoscalar $E0$ strength distributions deduced from the MDA analyses of
inelastic \aPb scattering data at 240 MeV by Youngblood \emph{et al.}
\cite{You04} and 386 MeV by Uchida \emph{et al.} \cite{Uch03,Uch03t} in
comparison with the RPA results. See details in text.

\begin{figure}
\begin{center}\hspace{-1cm}
\includegraphics[angle=0,scale=0.60]{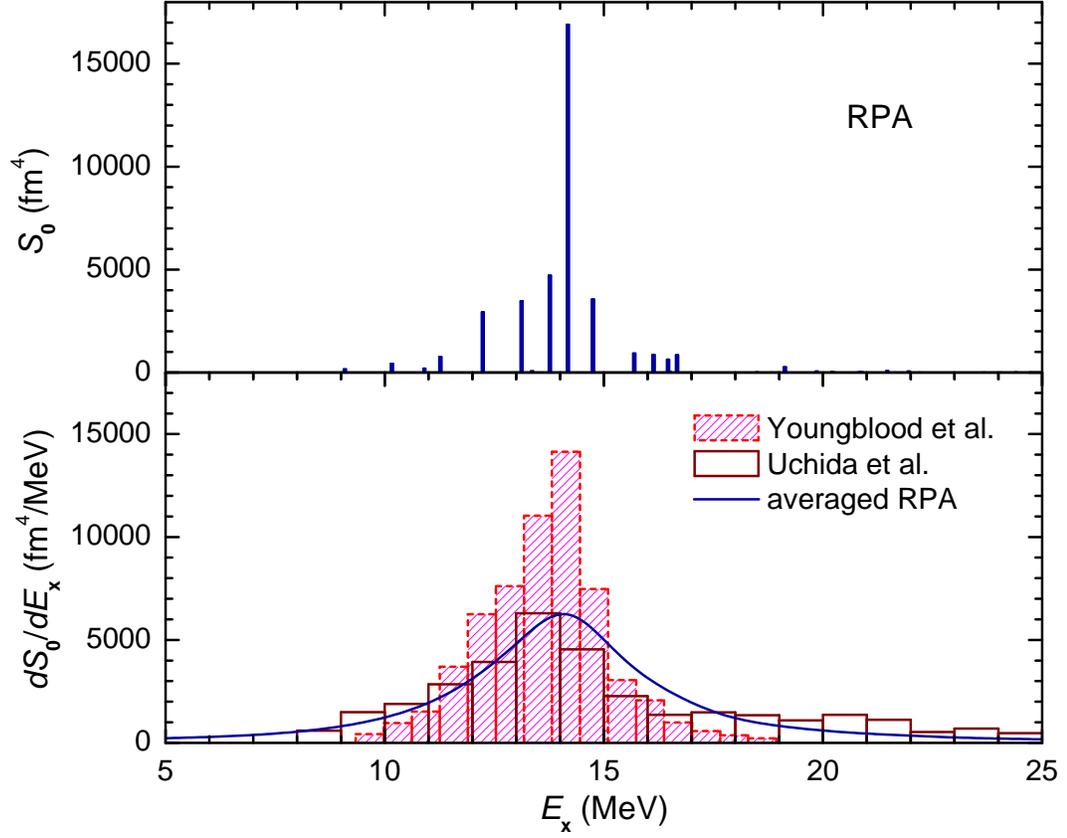}\vspace{-4cm}
\caption{Isoscalar $E0$ strength distributions deduced from the MDA analyses of
inelastic \aPb scattering data at 240 MeV by Youngblood \emph{et al.}
\cite{You04} and 386 MeV by Uchida \emph{et al.} \cite{Uch03,Uch03t} in
comparison with the RPA results. See details in text.}\label{f9}
\end{center}
\end{figure}

\begin{figure}[bht]
\begin{center}%\vspace{3cm}
\includegraphics[angle=0,scale=0.60]{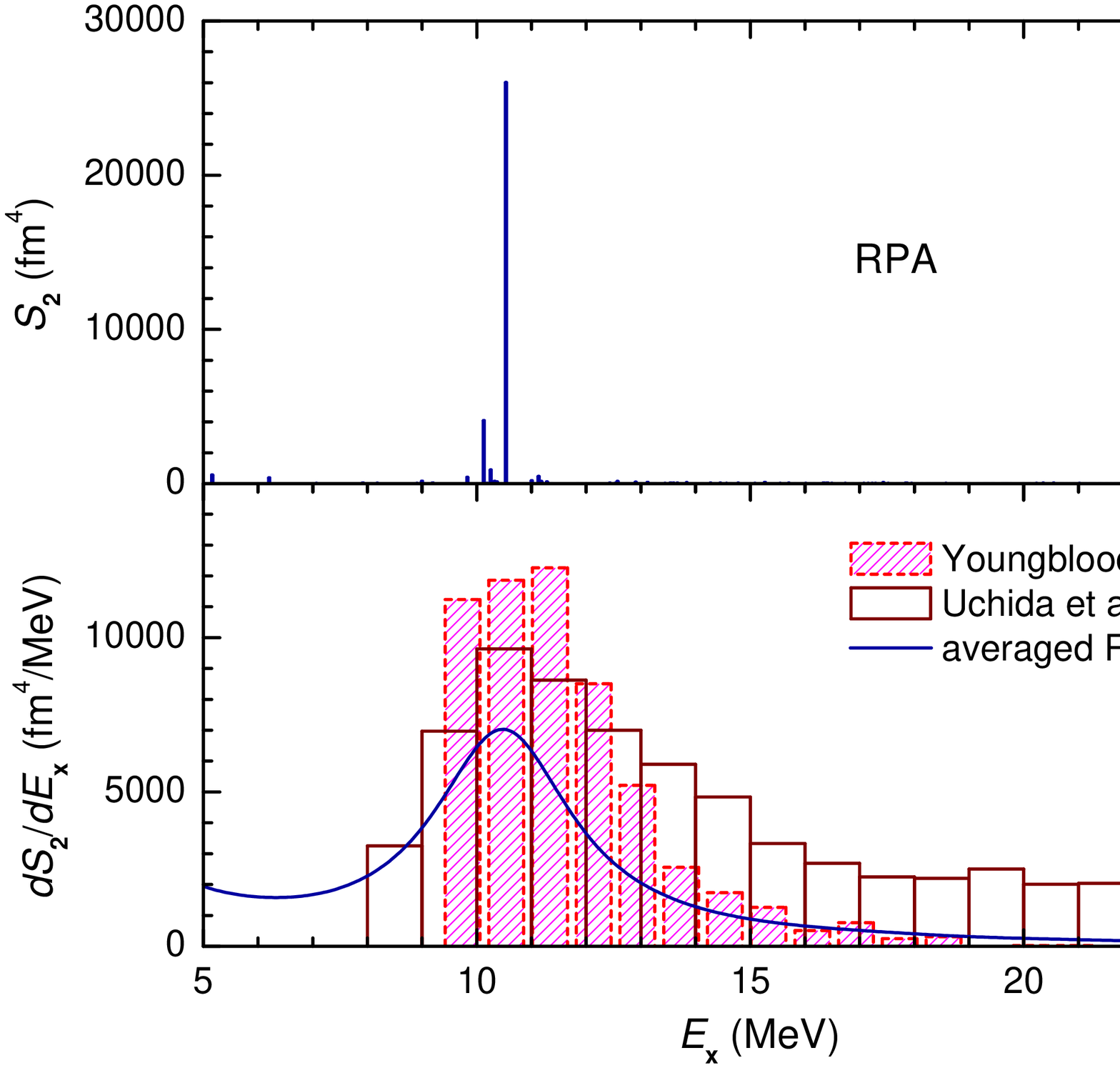}\vspace{-3.5cm}
\caption{The same as Fig.~\ref{f9} but for the isoscalar $E2$ strength
distributions.}\label{f10}
\end{center}
\end{figure}

\begin{figure}[bht]
\begin{center}%\vspace{3.5cm}
\includegraphics[angle=0,scale=0.60]{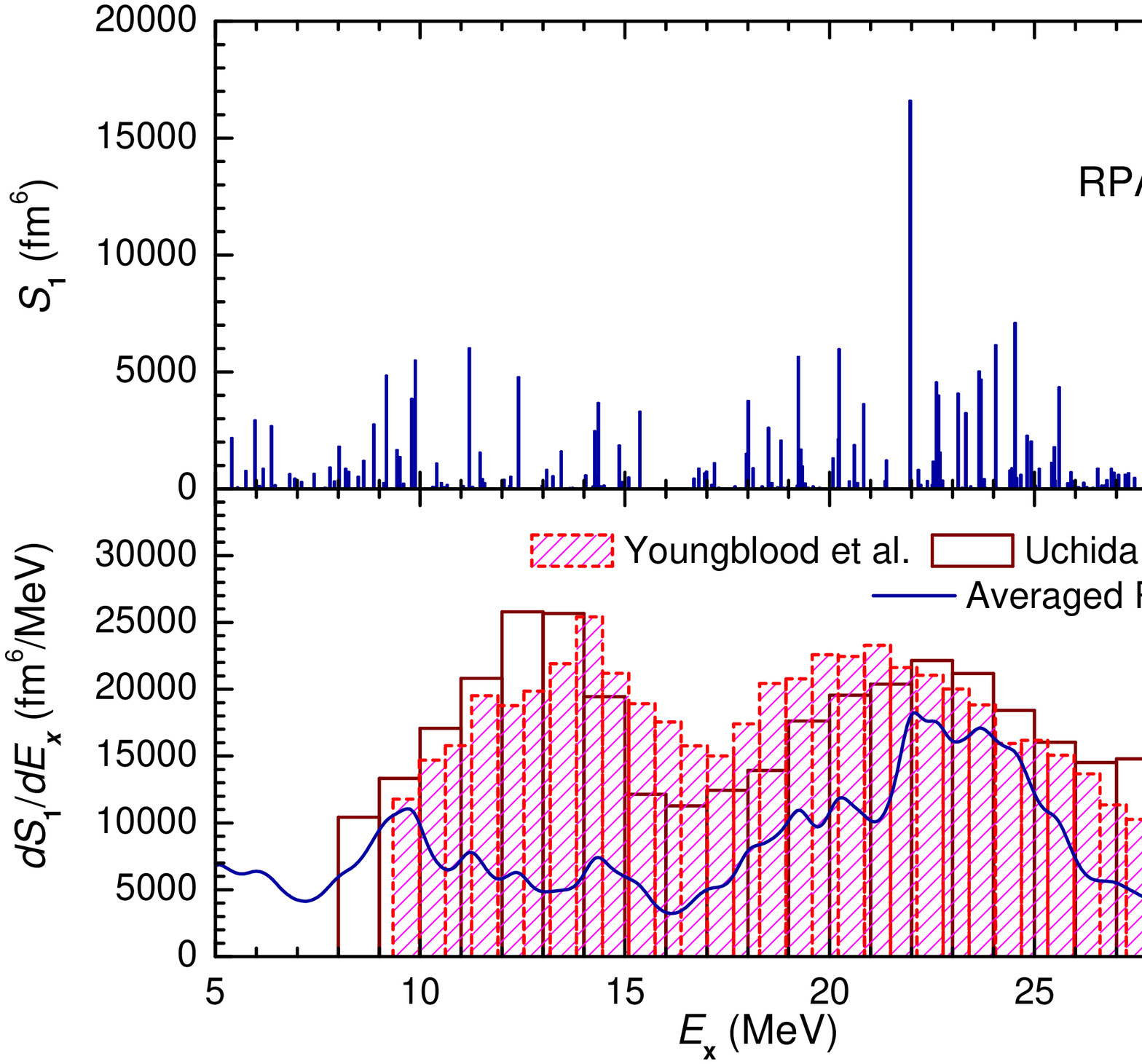}\vspace{-3.5cm}
\caption{The same as Fig.~\ref{f9} but for the isoscalar $E1$ strength
distributions.}\label{f11}
\end{center}
\end{figure}

\begin{figure}[bht]
\begin{center}\vspace{-2cm}
\includegraphics[angle=0,scale=0.65]{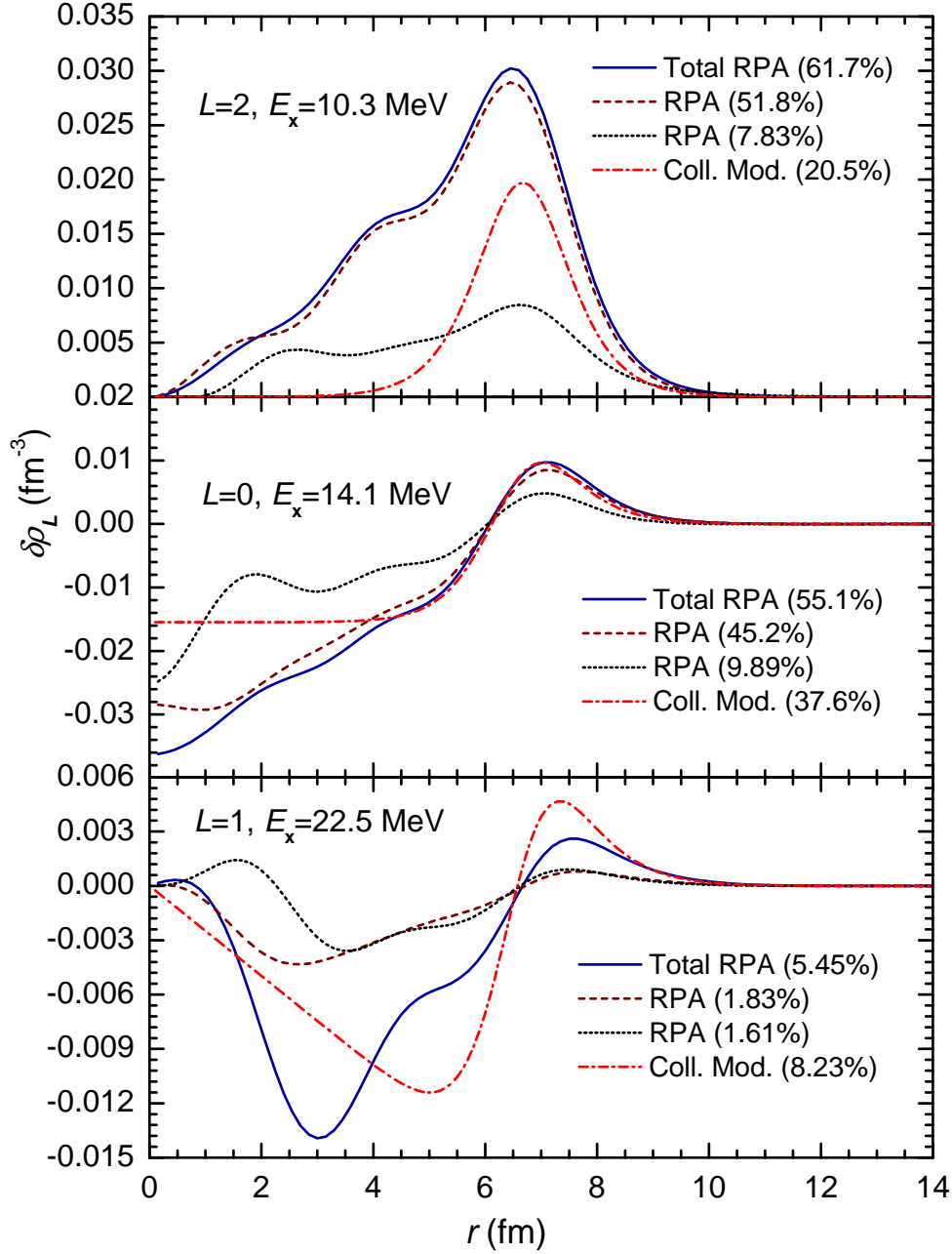}%\vspace{-3.5cm}
\caption{Total RPA transition density (\ref{denav1}) and transition densities of
the two strongest RPA states in the 640-keV energy bins centered at $E_x =
10.3$, 14.1 and 22.5 MeV, respectively. The corresponding collective model
transition densities were built upon the $EL$ strengths given by the MDA of 240
MeV data \cite{You04}. The quoted percentages are the exhausted fractions of the
isoscalar $EL$ EWSR.} \label{f12}
\end{center}
\end{figure}

\begin{figure}[bht]
 \begin{center}\vspace{-2cm}
\includegraphics[angle=0,scale=0.65]{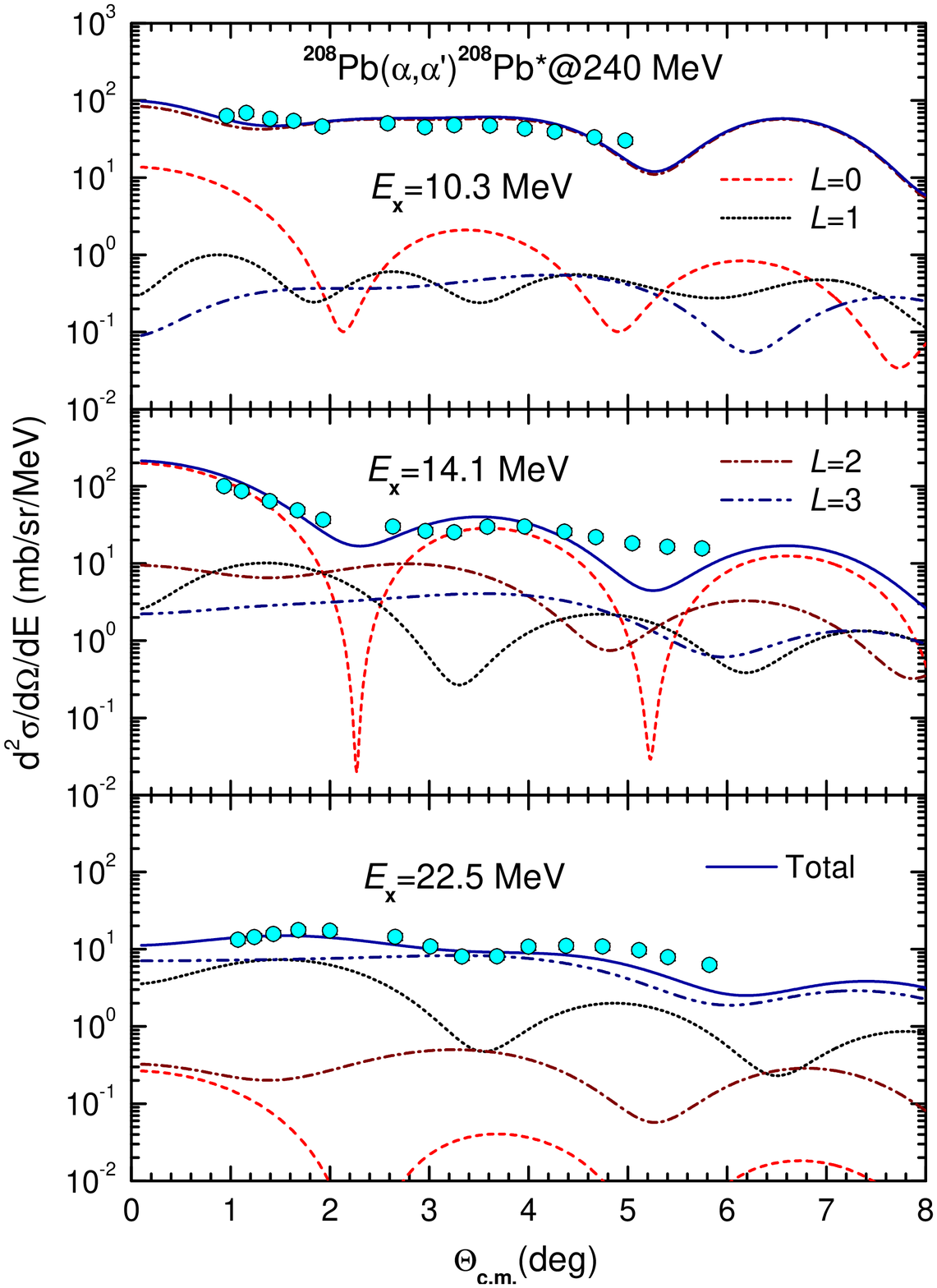}%\vspace{1cm}
\caption{Inelastic \aPb scattering data at $E_{\rm lab}=240$ MeV measured for
the 640 keV energy bins centered at $E_x = 10.3$, 14.1 and 22.5 MeV respectively
\cite{You04}, in comparison with the DFM + DWBA results obtained with the total
RPA transition densities (\ref{denav1}) which give the fractions of the
isoscalar $EL$ EWSR shown in Table~\ref{t3}.} \label{f13}
\end{center}
\end{figure}

Fig.11: The same as Fig.~\ref{f9} but for the isoscalar $E2$ strength
distributions.

Fig.12: The same as Fig.~\ref{f9} but for the isoscalar $E1$ strength
distributions.

Fig.13: Total RPA transition density (\ref{denav1}) and transition densities of
the two strongest RPA states in the 640-keV energy bins centered at $E_x =
10.3$, 14.1 and 22.5 MeV, respectively. The corresponding collective model
transition densities were built upon the $EL$ strengths given by the MDA of 240
MeV data \cite{You04}. The quoted percentages are the exhausted fractions of the
isoscalar $EL$ EWSR.

Fig.14: Inelastic \aPb scattering data at $E_{\rm lab}=240$ MeV measured for the
640 keV energy bins centered at $E_x = 10.3$, 14.1 and 22.5 MeV respectively
\cite{You04}, in comparison with the DFM + DWBA results obtained with the total
RPA transition densities (\ref{denav1}) which give the fractions of the
isoscalar $EL$ EWSR shown in Table~\ref{t3}. \pagebreak

\begin{table}\centering
\caption{Parameters of the complex density dependence of the CDM3Y6 interaction
(\ref{g1})-(\ref{g2}) used to calculate the OP and inelastic scattering FF for
the elastic and inelastic \aPb scattering at $E_{\rm lab}=240$ and 386 MeV.
$N_{\rm R(I)}$ are the renormalization coefficients of the real and imaginary OP
given by the optical model analysis of elastic scattering data; $\sigma_{\rm R}$
is the calculated total reaction cross section.} \vspace*{0.3cm}
\begin{tabular}{|c|c|c|c|c|l|c|c|c|} \hline
$E_{\rm lab}$ & $x$ & $C_x$ & $\alpha_x$ & $\beta_x$ & $\gamma_x$ &
$N_x$ & $\sigma_{\rm R}$  & $\sigma^{\rm exp}_{\rm R}$ \\
 (MeV) & & & & (fm$^3$) & (fm$^3$) & & (mb) & (mb) \\ \hline
 240 & R & 0.2243 & 3.8033 & 1.4099 & -4.0 & 0.9043 & 2768 &
 2900$\pm 190\ ^{a)}$\\
    & I & 0.1897 & 2.4840 & 5.1831  & -3.1341 & 1.4052 &  & \\
 386 & R & 0.1991 & 3.8033 & 1.4099 & -4.0 & 0.9885 & 2754 &
 2884$\pm 87\ ^{b)}$\\
    & I & 0.1435 & 3.1541 & 2.5646  & -2.5089 & 1.3565 & & \\ \hline
\end{tabular}\label{t1} \\
$^{a)}$ {\small Experimental total reaction cross section measured at
 $E_{\rm lab}=192$ MeV \cite{Ing00}} \\
$^{b)}$ {\small Experimental total reaction cross section measured at
 $E_{\rm lab}=340$ MeV \cite{Bon85}} \\
\end{table}

\begin{table}\vspace*{-2cm}\centering
\caption{Fractions of the isoscalar $EL$ EWSR exhausted in energy bin centered
at the excitation energy $E_x$ deduced from the MDA analyses of inelastic \aPb
scattering data at $E_{\rm lab}=240$ \cite{You04} and 386 MeV
\cite{Uch03,Uch04}, and from the \emph{averaged} RPA results. The averaging of
RPA transition densities ($<$RPA$>$) is discussed below in Sec.~\ref{sec4}.
$\delta_L$ are the deformation lengths for the CM nuclear transition densities
(\ref{fd6})-(\ref{fd11}) based on the $EL$ strengths taken from
Ref.~\cite{You04}.} \vspace*{0.3cm}
\begin{tabular}{|c|c|c|c|c|c|c|} \hline
\multicolumn{2}{|c|}{$E_x$} & $L^\pi$ & \multicolumn{3}{|c|}{$100\times
a_L(E_x)$ } &
 \multicolumn{1}{|c|}{$\delta_L$} \\
 \multicolumn{2}{|c|}{(MeV)} &  & \multicolumn{3}{|c|}{(\% EWSR/MeV)} &
 \multicolumn{1}{|c|}{(fm)}  \\ \hline
\cite{You04} & \cite{Uch03,Uch04} &  & MDA \cite{You04} & MDA
\cite{Uch03,Uch04} & $<$RPA$>$ &  MDA \cite{You04}\\
\hline
10.3 & 9.5 & $0^+$ & $1.88\pm 0.70$ & $2.68 \pm 0.47$ & 2.74 & 0.0086\\
     &     & $1^-$ & $2.63\pm 0.65$ & $2.20\pm 0.23$ & 1.39 & 0.0012 \\
     &     & $2^+$ & $20.5\pm 1.11$ & $12.4\pm 0.40$ & 13.3  & 0.2928 \\
     &     & $3^-$ & 0.0 & $3.48\pm 0.18$ & 1.06 & 0.0 \\ \hline
10.9 & 10.5& $0^+$ & $3.13\pm 0.70$ & $3.74\pm 0.65$ & 2.74 & 0.0111  \\
     &     & $1^-$ & $3.01\pm 0.50$  &$3.12\pm 0.45$ & 1.27 & 0.0014 \\
     &     & $2^+$ & $23.4\pm 1.51$  & $19.0\pm 0.80$ & 13.7 & 0.3094  \\
     &     & $3^-$ & $12.3\pm 0.20$ & $4.27\pm 0.46$ & 1.07 & 0.3354 \\ \hline
14.1 & 13.5& $0^+$&$37.6\pm 0.62 $ & $16.0\pm 1.86$& 16.6  & 0.0329 \\
     &     & $1^-$&$6.26\pm 0.42$ &$6.03\pm 0.78$& 1.68  & 0.0018 \\
     &     & $2^+$& $6.59\pm 1.79$ &$14.9\pm 0.80$& 3.12  & 0.1416 \\
     &     & $3^-$& 0.0 &$3.40\pm 0.76$& 1.43  & 0.0  \\ \hline
14.8 & 14.5& $0^+$&$20.8\pm 0.90$&$12.4\pm 1.60$& 16.2  & 0.0243 \\
     &     & $1^-$&$5.45\pm 0.32$&$4.90\pm 0.52$& 1.85  & 0.0016 \\
     &     & $2^+$ & $4.75\pm 2.26$ &$13.2\pm 1.40$& 2.73  & 0.1185 \\
     &     & $3^-$& 0.0 &$4.49\pm 0.68$& 1.48  & 0.0 \\ \hline
22.5 & 22.5& $0^+$& 0.0 &$2.22\pm 1.26$& 1.29 & 0.0 \\
     &     & $1^-$&$8.23\pm 0.15$&$8.67\pm 0.59$ & 7.06 & 0.0016 \\
     &     & $2^+$& 0.0 &$10.4\pm 1.70$& 0.60 & 0.0\\
     &     & $3^-$&$5.90\pm 0.30$&$5.42\pm 0.95$& 4.74 & 0.1584 \\ \hline
\end{tabular} \label{t2}
\end{table}

\begin{table}[htb]\centering \caption{Fractions of the isoscalar $EL$ EWSR
exhausted in the 640-keV energy bin centered at the excitation energy $E_x$
determined from the RPA transition strengths.} \vspace*{0.3cm}
\begin{tabular}{|c|c|c|c|} \hline
 \multicolumn{4}{|c|}{\% EWSR/MeV} \\ \hline
 $L^\pi$  & $E_x=10.3$ MeV & $E_x=14.1$ MeV & $E_x=22.5$ MeV \\ \hline
 $0^+$ &  1.25 & 55.1 & 0.06 \\
 $1^-$ &  0.32 & 2.25 & 5.45  \\
 $2^+$ &  61.7 & 0.34 & 0.06 \\
 $3^-$ &  0.69 & 0.83 & 10.8 \\ \hline
\end{tabular} \label{t3}
\end{table}

\begin{thebibliography}{00}
\bibitem{books} M.N. Harakeh, A. van der Woude, {\em Giant Resonances:
Fundamental High-Frequency Modes of Nuclear Excitation}, Clarendon Press,
Oxford, 2001; P.F. Bortignon, A. Bracco, R.A. Broglia, {\em Giant Resonances;
Nuclear Structure at Finite Temperature}, Harwood Academic, New York, 1998.
\bibitem{Shlo06} S. Shlomo, V.M. Kolomietz, G. Col\`o,
 Eur. Phys. J. A 30 (2006) 23-30.
\bibitem{Colo08} G. Col\`o, Physics of Particles and Nuclei
 39 (2008) 286-326.
\bibitem{Har77} M.N. Harakeh, K. van der Borg, T. Ishimatsu, H.P. Morsch,
 A. van der Woude, and F.E. Bertrand, Phys. Rev. Lett. 38 (1977) 676-679.
\bibitem{Har79} M.N. Harakeh, B. van Heyst, K. van der Borg,
 A. van der Woude, Nucl. Phys. A 327 (1979) 373-396.
\bibitem{You77} D.H. Youngblood, C.M. Rozsa, J.M. Moss, D.R. Brown,
 J.D. Bronson, Phys. Rev. Lett. 39 (1977) 1188-1191.
\bibitem{You04} D.H. Youngblood, Y.W. Lui, H.L. Clark,
 B. John, Y. Tokimoto, X. Chen, Phys. Rev. C 69 (2004) 034315; D.H.
 Youngblood, X. Chen, private communication (unpublished).
\bibitem{Uch03} M. Uchida, H. Sakaguchi, M. Itoh, M. Yosoi, T. Kawabata,
 H. Takeda, Y. Yasuda, T. Murakami, T. Ishikawa, T. Taki, N. Tsukahara, S.
Terashima, U. Garg, M. Hedden, B. Kharraja, M. Koss, B. K. Nayak, S. Zhu, M.
Fujiwara, H. Fujimura, K. Hara, E. Obayashi, H.P. Yoshida, H. Akimune, M.N.
Harakeh, M. Volkerts, Phys. Lett. B 557 (2003) 12-19.
\bibitem{Uch04} M. Uchida, H. Sakaguchi, M. Itoh, M. Yosoi, T. Kawabata,
Y. Yasuda, H. Takeda, T. Murakami, S. Terashima, S. Kishi, U. Garg, P.
Boutachkov, M. Hedden, B. Kharraja, M. Koss, B.K. Nayak, S. Zhu, M. Fujiwara, H.
Fujimura, H.P. Yoshida, K. Hara, H. Akimune, M.N. Harakeh,
 Phys. Rev. C 69 (2004) 051301(R); T. Kawabata, M. Uchida,
 private communication (unpublished).
\bibitem{Dieperink} M.N. Harakeh, A.E.L. Dieperink,
 Phys. Rev. C 23 (1981) 2329-2334.
\bibitem{giai} N.V. Giai, H. Sagawa, Nucl. Phys. A 371 (1981) 1-18.
\bibitem{Dav97} B.F. Davis, U. Garg, W. Reviol, M.N. Harakeh, A. Bacher,
 G.P. A. Berg, C.C. Foster, E.J. Stephenson, Y. Wang, J. J\"anecke, K. Pham,
 D. Roberts, H. Akimune, M. Fujiwara, J. Lisantti,
 Phys. Rev. Lett. 79 (1997) 609-612.
\bibitem{Sat83} G.R. Satchler, {\em Direct Nuclear Reactions},
 Clarendon Press, Oxford, 1983.
\bibitem{Kho00} D.T. Khoa, G.R. Satchler, Nucl. Phys. A 668 (2000) 3-41.
\bibitem{KhoSat97} G.R. Satchler, D.T. Khoa, Phys. Rev. C 55 (1997) 285-297.
\bibitem{Kho97} D.T. Khoa, G.R. Satchler, W. von Oertzen,
 Phys. Rev. C 56 (1997) 954-969.
\bibitem{Kho07} D.T. Khoa, W. von Oertzen, H.G. Bohlen, S. Ohkubo,
 J. Phys. G 34 (2007) R111-R164.
\bibitem{Je77} J.P. Jeukenne, A. Lejeune, C. Mahaux,
 Phys. Rev. C 16 (1977) 80-96.
\bibitem{Kho08} D.T. Khoa, D.C. Cuong, Phys. Lett. B 660 (2008) 331-338.
\bibitem{An83} N. Anantaraman, H. Toki, G.F. Bertsch,
 Nucl. Phys. A 398 (1983) 269-278.
\bibitem{Ing00} A. Ingemarsson, J. Nyberg, P.U. Renberg, O. Sundberg,
 R.F. Carlson, A.J. Cox, A. Auce, R. Johansson, G. Tibell, D.T. Khoa,
 R. E. Warner, Nucl. Phys. A 676 (2000) 3-31.
\bibitem{Bon85} B. Bonin, N. Alamanos, B. Berthier, G. Bruge, H. Faraggi,
 J.C. Lugol, W. Mittig, L. Papineau, A.I. Yavin, J. Arvieux, L. Farvacque,
 M. Buenerd, W. Bauhoff, Nucl. Phys. A 445 (1985) 381-407.
\bibitem{Kho01} D.T. Khoa, Phys. Rev. C 63 (2001) 034007.
\bibitem{Lo78} W.G. Love, Nucl. Phys. A 312 (1978) 160-176.
\bibitem{Car96} F. Carstoiu, M. Lassaut, Nucl. Phys. A 597 (1996) 269-297.
\bibitem{Hor95} D.J. Horen, J.R. Beene, G.R. Satchler,
 Phys. Rev. C 52 (1995) 1554-1564.
\bibitem{Bohr} A. Bohr, B.R. Mottelson, {\em Nuclear Structure},
 W.A. Benjamin, Reading, 1975, vol. II.
\bibitem{Far85} M.E. Farid, G.R. Satchler, Nucl. Phys. A 438 (1985) 525-535.
\bibitem{Uberall} H. \"Uberall, {\em Electron Scattering from
Complex Nuclei}, Academic Press, New York, 1971, vol. B.
\bibitem{Woude99} A. van der Woude, Nucl. Phys. A 649 (1999) 97c-103c.
\bibitem{khoapp} D.T. Khoa, E. Khan, G. Col\`o, N.V. Giai,
 Nucl. Phys. A 706 (2002) 61-84.
\bibitem{Guillot} J. Guillot, S. Gal\`es, D. Beaumel, S. Fortier,
E. Rich, N. Van Giai, G. Col\`o, A.M. van der Berg, S. Brandenburg, B. Davids,
M.N. Harakeh, M. Hunyadi, M. de Huu, S.Y. van der Werf, H.J. W\"ortche, C.
B\"aumer, D. Frekers, E.W. Grewe, P. Haefner, B.C. Junk, M. Fujiwara,
 Phys. Rev. C 73 (2006) 014616.
\bibitem{Chabanat} E. Chabanat, P. Bonche, P. Haensel,
 J. Meyer, R. Schaeffer, Nucl. Phys. A 635 (1998) 231-256.
\bibitem{comex2} G. Col\`o, P.F. Bortignon, S. Fracasso, N. V. Giai,
 Nucl. Phys. A 788 (2007) 137c-181c.
\bibitem{Sa79} G.R. Satchler, W.G. Love, Phys. Rep. 55 (1979) 183-254.
\bibitem{Kib02} T. Kibedi, R.H. Spear,
 At. Data and Nucl. Data Tables 80 (2002) 35-82.
\bibitem{Col04} G. Col\`o, N. Van Giai, J. Meyer, K. Bennaceur,
 P. Bonche, Phys. Rev. C 70 (2004) 024307.
\bibitem{Vre03} D. Vretenar, T. Nik\v{s}i\'{c}, P. Ring,
 Phys. Rev. C 68 (2003) 024310.
\bibitem{Uch03t} M. Uchida, PhD Thesis (University of Kyoto, 2003).
%\bibitem{Sag08} H. Sagawa, private communication (unpublished).
\bibitem{Str82}S. Stringari, Phys. Lett. 108 B (1982) 232-236.
\bibitem{Bla95} J.P. Blaizot, J.F. Berger, J. Decharg\'e, M. Girod,
 Nucl. Phys. A 591 (1995) 435-457.
\bibitem{Kho86} D.T. Khoa, I.N. Kuchtina, V.Yu. Ponomarev,
 Sov. J. Nucl. Phys. 44 (1986) 585-590.
\bibitem{Ho96} D.J. Horen, G.R. Satchler, S.A. Fayans, E.L. Trykov,
 Nucl. Phys. A 600 (1996) 193-235.
\bibitem{Kol00} A. Kolomiets, O. Pochivalov, S. Shlomo,
 Phys. Rev. C 61 (2000) 034312.
\bibitem{Fe92} H. Feshbach, {\sl Theoretical Nuclear Physics -
 Nuclear Reactions}, Wiley, New York, 1992.
\bibitem{Col00} G. Col\`o, N.V. Giai, P.F. Bortignon, M.R. Quaglia,
 Phys. Lett. B 485 (2000) 362-366.
\bibitem{Vre00} D. Vretenar, A. Wandelt, P. Ring,
 Phys. Lett. B 487 (2000) 334-340.
\bibitem{Pie01} J. Piekarewicz, Phys. Rev. C 64 (2001) 024307.
\bibitem{toroidal} S.I. Bastrukov, S. Misicu, V.I. Sushkov,
 Nucl. Phys. A 562 (1993) 191; S. Misicu, Phys. Rev. C 73 (2006) 024301;
 D. Vretenar, N. Paar, P. Ring, T. Nik\v{s}i\'{c},
 Phys. Rev. C 65 (2002) 021301(R).
\bibitem{Ber83} G.F. Bertsch, P.F. Bortignon, R.A. Broglia,
 Rev. Mod. Phys. 55 (1983) 287-314.
\bibitem{Wam90} S. Drozdz, S. Nishizaki, J. Speth, J. Wambach,
 Phys. Rep. 197 (1990) 1-65.
\bibitem{Ber80} F.E. Bertrand, G.R. Satchler, D.J. Horen, J.R. Wu, A.D. Bacher,
 G.T. Emery, W.P. Jones, D.W. Miller, A. van der Woude,
 Phys. Rev. C 22 (1980) 1832-1847.
\bibitem{Kam09} V. Tselyaev, J. Speth, S. Krewald, E. Litvinova, S.
 Kamerdzhiev, N. Lyutorovich, A. Adveenkov, F. Gr\"ummer,
 Phys. Rev. C79 (2009) 034309.
\end{thebibliography}
\end{document}